\documentclass[review]{elsarticle}
\usepackage{verbatim}
\usepackage{lineno,hyperref}
\modulolinenumbers[5]

\begin{document}

\begin{frontmatter}

\title{\Large Coherent pion production in neutrino (anti-neutrino)-nucleus
  interaction}

\author{\large Hariom Sogarwal}
\author{\large Prashant Shukla\corref{mycorrespondingauthor}}
\cortext[mycorrespondingauthor]{Corresponding author}
\ead{pshuklabarc@gmail.com}
\address{Nuclear Physics Division, Bhabha Atomic Research Centre, Mumbai 400085,
  India}
\address{Homi Bhabha National Institute, Anushakti Nagar, Mumbai 400094,
  India}

\begin{abstract}
We present a study of coherent pion production in neutrino-nucleus interactions
using the formalism based on partially
conserved axial current theorem which connects the
neutrino-nucleus cross section to the pion-nucleus elastic scattering cross section.
Pion-nucleus elastic scattering cross section is calculated using Glauber
model which takes three inputs, nuclear densities, pion-nucleon cross section and
$\alpha_{\pi N}$ which is the ratio of real to imaginary part of $\pi N$ forward
scattering amplitude, 
for which the parametrizations are obtained from measured data.
We calculate the differential and integrated cross sections for charge and
neutral current coherent pion production in neutrino (anti-neutrino)-nucleus
scattering for a range of nuclear targets from light to heavy materials such as lithium, 
carbon, hydrocarbon, oxygen, silicon, argon, iron and lead.
The results of these cross section calculations are
compared with the measured data and with the calculations from the Berger-Sehgal
model and GENIE package.
There is an excellent agreement
between the calculated and measured cross sections with Glauber
model. While GENIE and Berger-Sehgal model give a good description
of the data in the lower energy range the
present calculations describe the data in all energy ranges.
 Predictions are also made for upcoming experiments like INO
 and DUNE in the coherent region of neutrino cross section.
\end{abstract}


\end{frontmatter}


\section{Introduction}

 The neutrinos produced in upper atmosphere and in 
accelerators are used for studying the phenomena of neutrino oscillations
by many experiments worldwide~\cite{Messier:2006yg,Diwan:2016gmz,Zyla:2020zbs}.
  Most experiments focussing on muon neutrinos are designed to measure recoil muons
when neutrinos undergo charge current interaction in the detector
medium. Some examples of detector media are; iron in case of India-based 
Neutrino Observatory (INO)~\cite{Ahmed:2015jtv} and argon in case of
DUNE experiment~\cite{Abi:2020qib}.
Bulk of the interactions in the detector comes from intermediate energy neutrinos
having contribution from many processes which include quasi elastic scattering,
interaction via resonance pion production and deep inelastic
scattering\,\cite{Formaggio:2013kya,Saraswat:2016tqa,Grover:2018ggi}.
One of the important processes in the resonance production region is
coherent pion production in which the nucleus interacts as a whole
with the neutrino with  its quantum state remains unchanged.
 The charge current (CC) and neutral current (NC) coherent pion production processes are given as 
\begin{eqnarray*}
\nu_{\mu} + A  & \rightarrow  \mu^{-} + \pi^{+} + A. \,\,\, {\rm (CC)} \\
\nu_{\mu} + A  & \rightarrow  \nu_{\mu} + \pi^{0} + A~. \,\,\, {\rm (NC)}
\label{cccoherentpionproductionequation}
\end{eqnarray*}
The four-momentum transfer $Q$ between the incoming neutrino and outgoing lepton
is given by $Q^{2}=-q^{2}={\bf q}^{2}-\nu^{2}$. 
Here, ${\bf q}$ is the 3-momentum transfer and $\nu~(=E_{\nu}-E_{\mu})$ is the energy difference
between the incident neutrino (with energy $E_{\nu}$) and outgoing
lepton (with energy $E_{\mu}$).
For coherent pion production, the squared momentum transfer to the nucleus from
the lepton-pion system $|t|=|(q-p_{\pi})^{2}|$ remains small.
Here $p_{\pi}$ is the 4-momentum of outgoing pion.
  Estimating coherent pion production is important for the analysis of data of
neutrino oscillation experiments.
The simplest theoretical  approach for describing coherent pion production
is based on Adler's Partially Conserved Axial Current (PCAC) theorem which
relates the neutrino induced coherent pion production to the pion-nucleus
elastic scattering~\cite{Rein:1982pf, Kopeliovich:1992ym, Kopeliovich:2004px, Paschos:2005km, Berger:2008xs, Paschos:2009ag}.
The PCAC model has been successful in describing coherent pion production at
high energy~\cite{Rein:1982pf}. Work with the same assumption
has been used at low energy in Ref.~\cite{Paschos:2005km}. There are several
microscopic models as well for coherent scattering e.g. in
Refs.~\cite{Singh:2006bm, AlvarezRuso:2007tt, AlvarezRuso:2007it, Amaro:2008hd}.
The GENIE (\textbf{G}enerates \textbf{E}vents for
\textbf{N}eutrino \textbf{I}nteraction \textbf{E}xperiments)~\cite{GENIEref}
uses PCAC theorem with Rein-Sehgal model~\cite{Rein:1982pf} for the coherent
neutrino-nucleus scattering which results in the production of forward going
pions in both charge current and neutral current interactions.

To obtain the elastic pion-nucleus scattering cross section, the Berger-Sehgal (BS)
model~\cite{Berger:2008xs} is used in literature which basically follows 
Rein-Sehgal (RS) method~\cite{Rein:1982pf} with some improvement in the way
parametrizations are done for elementary cross sections. 
In this model, the elastic scattering pion nucleus cross section is obtained from
measured total and elastic pion nucleon cross sections.
Further an exponential function is assumed for $t$ dependence of elastic scattering
cross section and attenuation of pion is made dependent on nuclear size.
GENIE uses RS model and its difference with BS calculations can arise due to
input elementary cross sections and other input parameters. 
The work presented in Ref.~\cite{Saraswat:2016kln} calculates the 
pion-nucleus elastic scattering cross section using the Glauber model in terms of measured
nuclear densities and measured pion nucleon cross sections.
In view of the recent data at various energies we aim to revisit this
model and improve the treatment of input parameters namely
total pion nucleon cross sections, $\alpha_{\pi N}$ (the ratio of real to
imaginary part of forward scattering amplitude) and
nuclear density function parameters. 
The parametrizations of the total pion nucleon cross sections 
have been done and presented using the latest data.
 The values of $\alpha_{\pi N}$ obtained with the elementary scattering amplitudes
may change when one applies the formalism to nuclear targets and are typically
around one \cite{Mehndiratta:2017gry}. In view of this, we use the pion nucleus
elastic scattering differential cross section data available for few incident
energies of pions on a broad range of nuclear targets to obtain $\alpha_{\pi N}$.
We also compare with the Berger-Sehgal model which gives reasonable description
of elastic scattering differential cross section at lower scattering angles only. 
 Recently, charge and neutral current coherent pion production has been measured for several
nuclei by many experiments such as MINER$\nu$A~\cite{Mislivec:2017qfz,tminervadata}, CHARM-II~\cite{Vilain:1993sf}, Aachen-Padova~\citep{faissner:1983}, Gargamelle~\cite{isiksal:1984}, CHARM~\cite{Bergsma:1985qy}, SKAT~\cite{Grabosch:1985mt},
NOMAD~\cite{Kullenberg:2009pu}, 15'B.C.~\cite{Baltay:1986cv},
SciBooNE~\cite{Kurimoto:2009wq}, MINOS~\cite{Adamson:2016hyz}, NO$\nu$A~\cite{Acero:2019qcr}, K2K~\cite{Ahn:2006zza}, T2K\cite{t2kflux, T2Kdata} and MiniBooNE~\cite{miniboonedata}.
The comparison of Glauber approach has been done with
the calculations from Berger-Sehgal (BS) model and the GENIE package (version v3\_00\_06a).

 In this work, we calculate coherent pion production in neutrino-nucleus interactions
in the resonance region using the PCAC based formalism.
Pion-nucleus elastic scattering cross section is calculated with the Glauber
model using pion-nucleon cross section and $\alpha_{\pi N}$.
The parametrizations of latest measured total pion nucleon cross section has been
presented. 
We calculate the differential and integrated cross sections for charge and
neutral current coherent pion production in neutrino (anti-neutrino)-nucleus
scattering for a range of nuclear targets: lithium, carbon, hydrocarbon, scintillator,
oxygen, silicon, argon, iron, and lead using the model with fixed parameters.
The results of these cross section calculations are compared with the measured
data, BS model and GENIE package.

\section{The formulation of the model}

The differential cross section for the charge current coherent pion production
scattering process~\cite{Kopeliovich:1992ym, Berger:2008xs} is 
\begin{eqnarray}
  \frac{d\sigma^{CC}}{dQ^{2} d\nu d|t|} &=& \frac{G^{2}_{F} \cos^{2}\theta_{C} f^{2}_{\pi}}{2 \pi^{2}}
   \frac{u v}{|{\bf q}|}~
   \Bigg[\Big(G_{A}-\frac{1}{2} \frac{Q^{2}_{m}} {(Q^{2}+m^{2}_{\pi})}\Big)^{2}  \nonumber  \\
&~& + \frac{\nu}{4 E_{\nu}} (Q^{2}-Q^{2}_{m}) \frac{Q^{2}_{m}}  {(Q^{2}+m^{2}_{\pi})^{2}}\Bigg]   
    \times \frac{d\sigma(\pi A \rightarrow \pi A)}{d|t|}~.
\label{marage1993}
\end{eqnarray}
For the neutral current, the above expression is modified as
\begin{eqnarray}
\frac{d\sigma^{NC}}{dQ^{2} d\nu d|t|} &=& \frac{G^{2}_{F} f^{2}_{\pi}}{4 \pi^{2}} 
\frac{G^{2}_{A}}{|{\bf q}|}~u~v~
\times \frac{d\sigma(\pi A \rightarrow \pi A)}{d|t|}~.
\label{nceqlabel}
\end{eqnarray}
Here $G_{F}$ (=1.16639 $\times 10^{-5}$ GeV$^{-2}$) is the Fermi coupling constant
and $\cos\theta_{C}~(=0.9725)$.
The kinematic factors $u$ and $v$
are given by : $u,v=\Big(E_{\nu}+E_{\mu}~\pm~|{\bf q}|\Big)/(2~E_{\nu})$.
The pion decay constant is $f_{\pi}~(=0.93 ~m_{\pi})$ and
${d\sigma(\pi A \rightarrow \pi A)}/{d|t|}$ is the pion-nucleus differential
elastic cross section calculated assuming $\nu$ as the total incident energy of
pion in laboratory frame. 
The axial vector form factor can be defined as
$G_{A}=m^{2}_{A}/(Q^{2}+m^{2}_{A})$ ~\cite{Berger:2008xs} with the axial
vector meson mass $m_{A}$ (= 1.05 GeV/c$^{2}$)~\cite{Saraswat:2016tqa}.
The high energy approximation to the true minimal
$Q^{2}$ is given by $Q^{2}_{m}=m^{2}_{l}~\nu/(E_{\nu}-\nu)$, where $m_{l}$ is the mass of
outgoing lepton which will be muon mass in case of CC.
 The expression for neutral current (NC) (Eq.\,\ref{nceqlabel}) is obtained
from Eq.\,\ref{marage1993} by putting $m_{l}$=0, $\theta_{C}$=0 and divide the
right hand side of the Eq.\,\ref{marage1993} by 2, because
$f_{\pi^{0}} = f_{\pi}/\sqrt{2}$.

The kinematic limits are guided by the work in Ref.~\cite{Paschos:2005km}.
The upper limit of $Q^2$ is taken as 1.0 GeV$^2$.
The $\nu$ integration should be done in the range
$\rm max (\xi \sqrt{Q^{2}} , \nu_{min} ) <   \nu   <  \nu_{max}$.
In the previous work~\cite{Saraswat:2016kln} the calculations were performed at two
values of $\xi$ (=1,2). In the present work, we include this 
variation as uncertainty in both Glauber and BS calculations. \\
\ \\
{\bf\large Pion-nucleus differential cross section} \\

To obtain the elastic pion-nucleus scattering cross section, the Berger-Sehgal (BS)
model~\cite{Berger:2008xs} is used in literature which basically follows 
Rein-Sehgal (RS) method~\cite{Rein:1982pf}. In their model, the pion-nucleus elastic scattering
cross section is obtained from measured total pion nucleon cross sections.
Further an exponential function is assumed for $t$ dependence and the attenuation
of pions is made dependent on nuclear radius. 
 We use scattering theory to obtain the pion-nucleus differential elastic cross section
given as 
\begin{equation}
\frac{d\sigma_{el}}{d|t|} = \frac{\pi}{k^{2}}|f(t)|^{2}, 
\label{dsigmadtpresent}
\end{equation}
where $f(t)$ is given by
\begin{eqnarray}
  f(t) = \frac{1}{2 i k} \sum\limits_{l=0}^{\infty} \Big(2 l + 1\Big)
  \Big(S_{l}-1\Big) 
P_{l}\Big(\cos\theta\Big) .
\label{ftdsigmaeldtterm}
\end{eqnarray}
Here $t = - 4k^{2} \, \sin^{2}\theta/2$ and $k$ is the momentum of pion and $\theta$
is the scattering angle of pion in center of mass frame.

We use the Glauber model to obtain the scattering matrix $S_{l}$ in terms of
pion-nucleus impact parameter $b$ 
by~\cite{Shukla:2001mb}
\begin{equation}
S_{l} = \exp(i\chi(b)), \quad b k = \Big(l + \frac{1}{2}\Big),~
\end{equation}
The Glauber phase shift $\chi(b)$ is given by  

\begin{equation}
\chi(b) = \frac{1}{2}\sigma_{\pi N} \Big(\alpha_{\pi N} + i\Big) 
A T(b)
\end{equation}
Here $\sigma_{\pi N}$ is the average total pion-nucleon cross section and
$\alpha_{\pi N}$ is the ratio of real to imaginary part of the
$\pi N$ forward scattering amplitude.
$T(b)$ is the overlap function obtained by the Glauber model, the full details of which are
given in Ref.~\cite{Saraswat:2016kln}.
The pion-nucleon cross sections are taken from the
Particle Data Group~\cite{Zyla:2020zbs}.

\begin{figure*}
\includegraphics[width=0.496\linewidth]{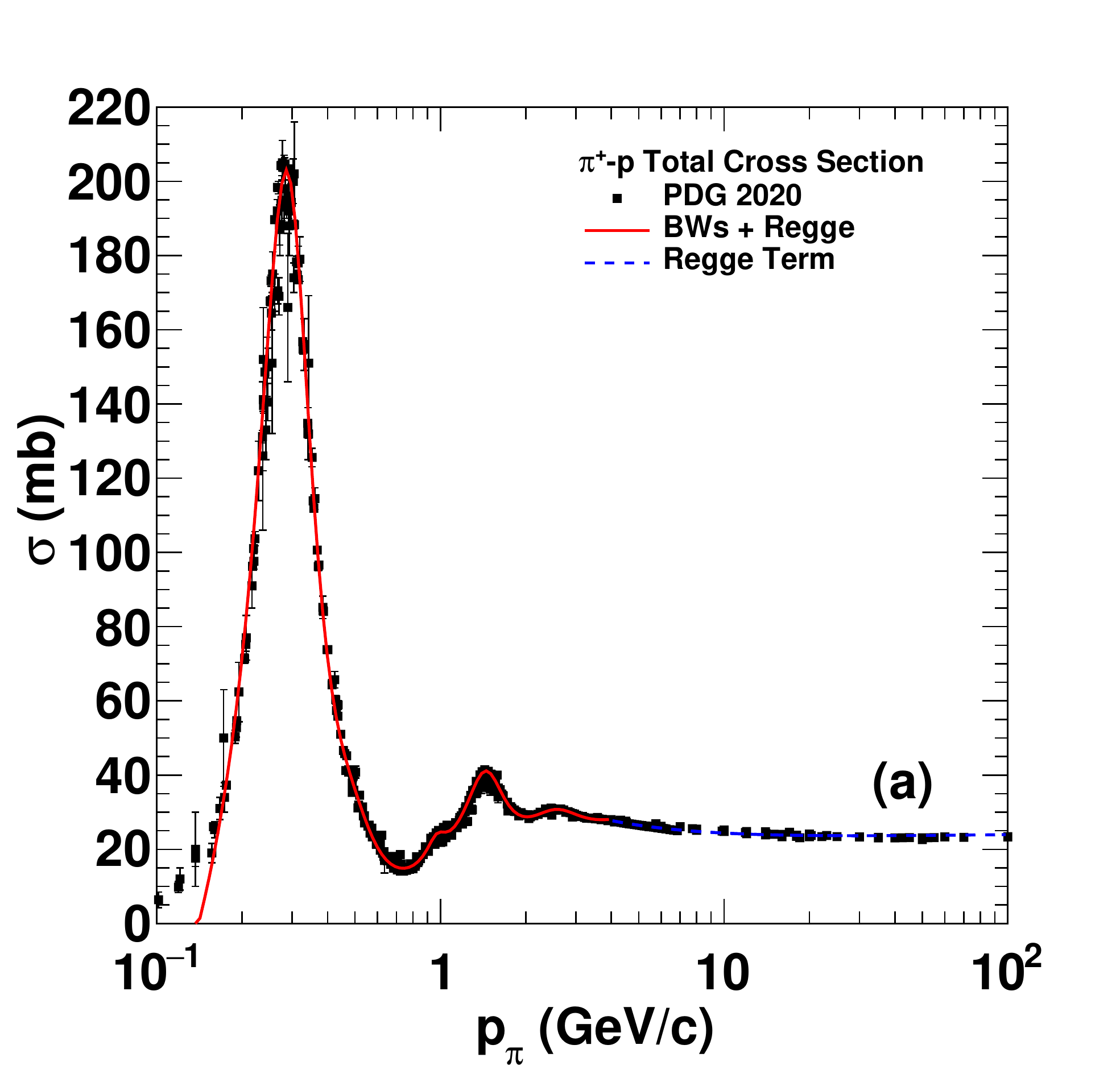}
\includegraphics[width=0.496\linewidth]{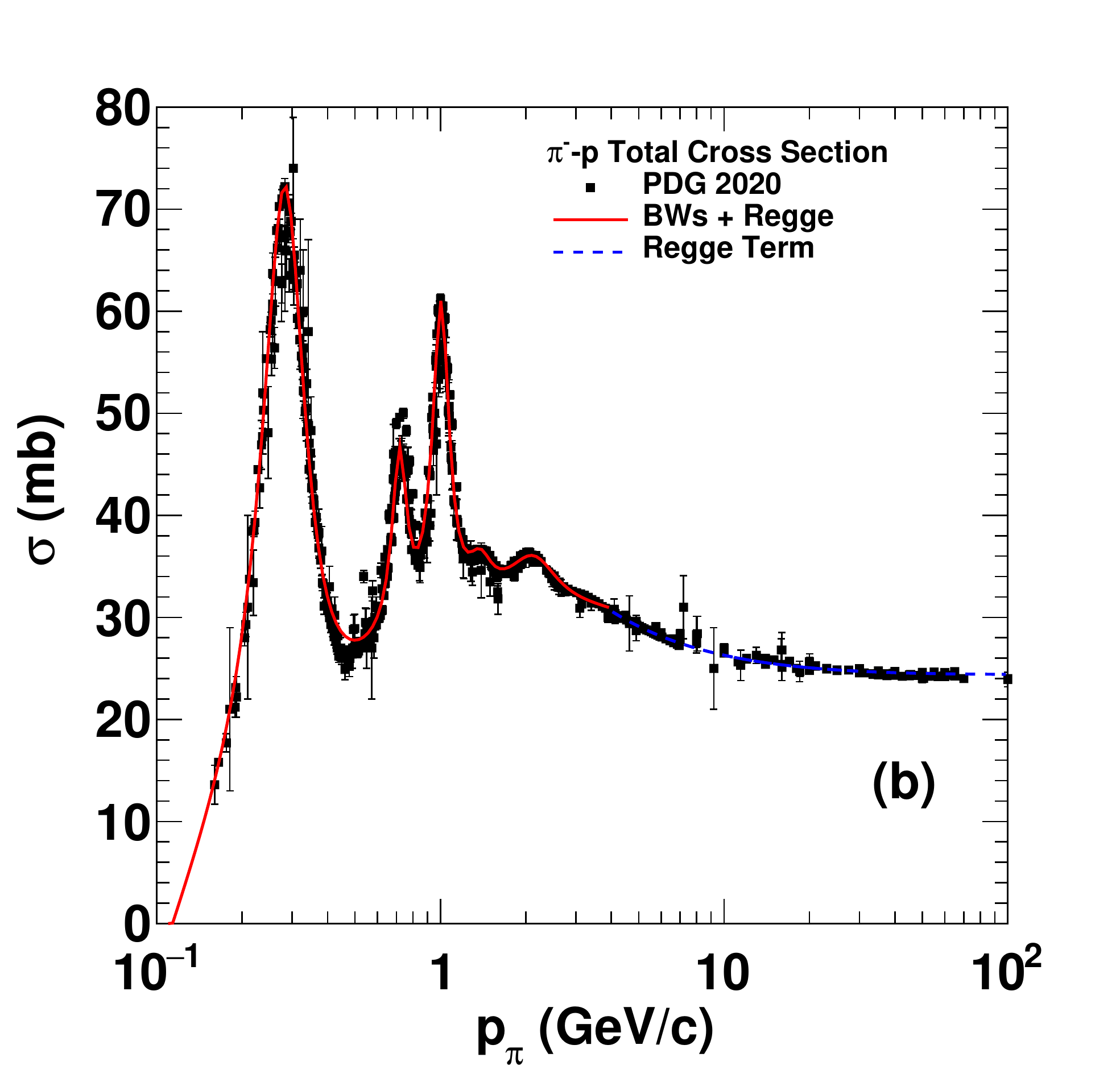}
\caption{Measured total cross sections for (a) $\pi^{+}$p and (b) $\pi^{-}$p collisions
as a function of pion momentum along with the fit function.}
\label{piminusptotal}
\end{figure*}

Figure\,\ref{piminusptotal} (a) and (b) show the measured total cross sections
for $\pi^{+}$p and $\pi^{-}$p collisions respectively as a function of pion momentum in laboratory frame.
The data is fitted with Breit-Wigners and Regge function for pion momentum up to
4 GeV/$c$ and a pure Regge
function above 4 GeV/$c$. The Regge term above 4 GeV/$c$ is given by 
$25.06 - 15.16/\sqrt{p_{\pi}} + 41.07\sqrt{p_{\pi}}$ for $\pi^{+}$p case and 
$24.73 -6.91/\sqrt{p_{\pi}} + 37.37\sqrt{p_{\pi}}$ for $\pi^{-}$p case.

\begin{figure*}
\begin{center}
\includegraphics[width=0.65\linewidth]{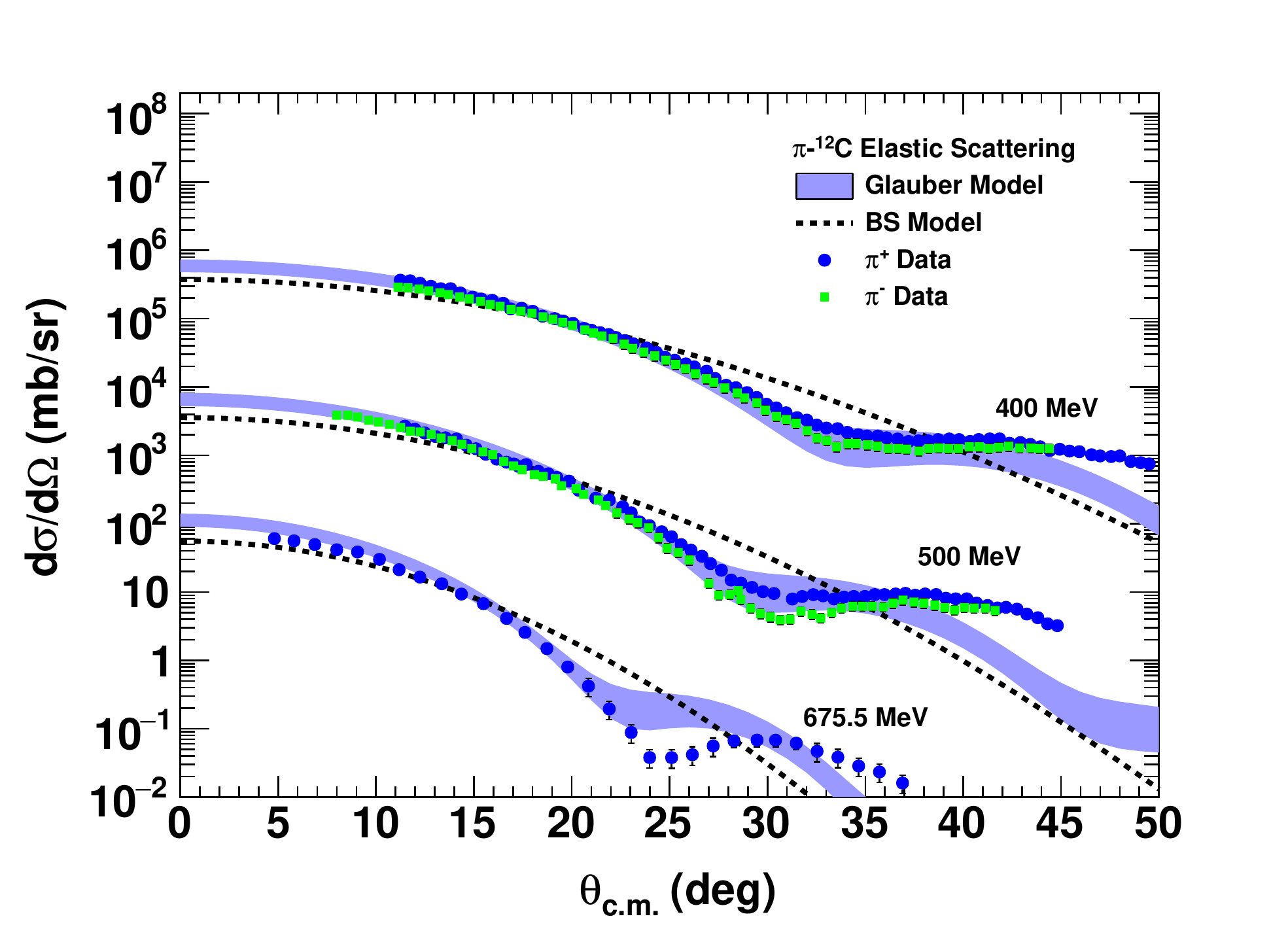}
\caption{Pion-$^{12}$C elastic scattering differential cross section as function of
pion scattering angle at three pion kinetic energies of 400 MeV, 500 MeV and
675.5 MeV scaled respectively by factors 1000, 10 and 0.1~\cite{Kahrimanis:1997if,BNL}.
The blue bands are obtained using the Glauber model for $\alpha_{\pi N}$ = 0.9-1.4 and
dashed lines are the calculations of the Berger-Sehgal (BS) model~\cite{Berger:2008xs}.}
\label{ElasticPionCarbon}
\end{center}
\end{figure*}

\begin{figure*}
\begin{center}
\includegraphics[width=0.65\linewidth]{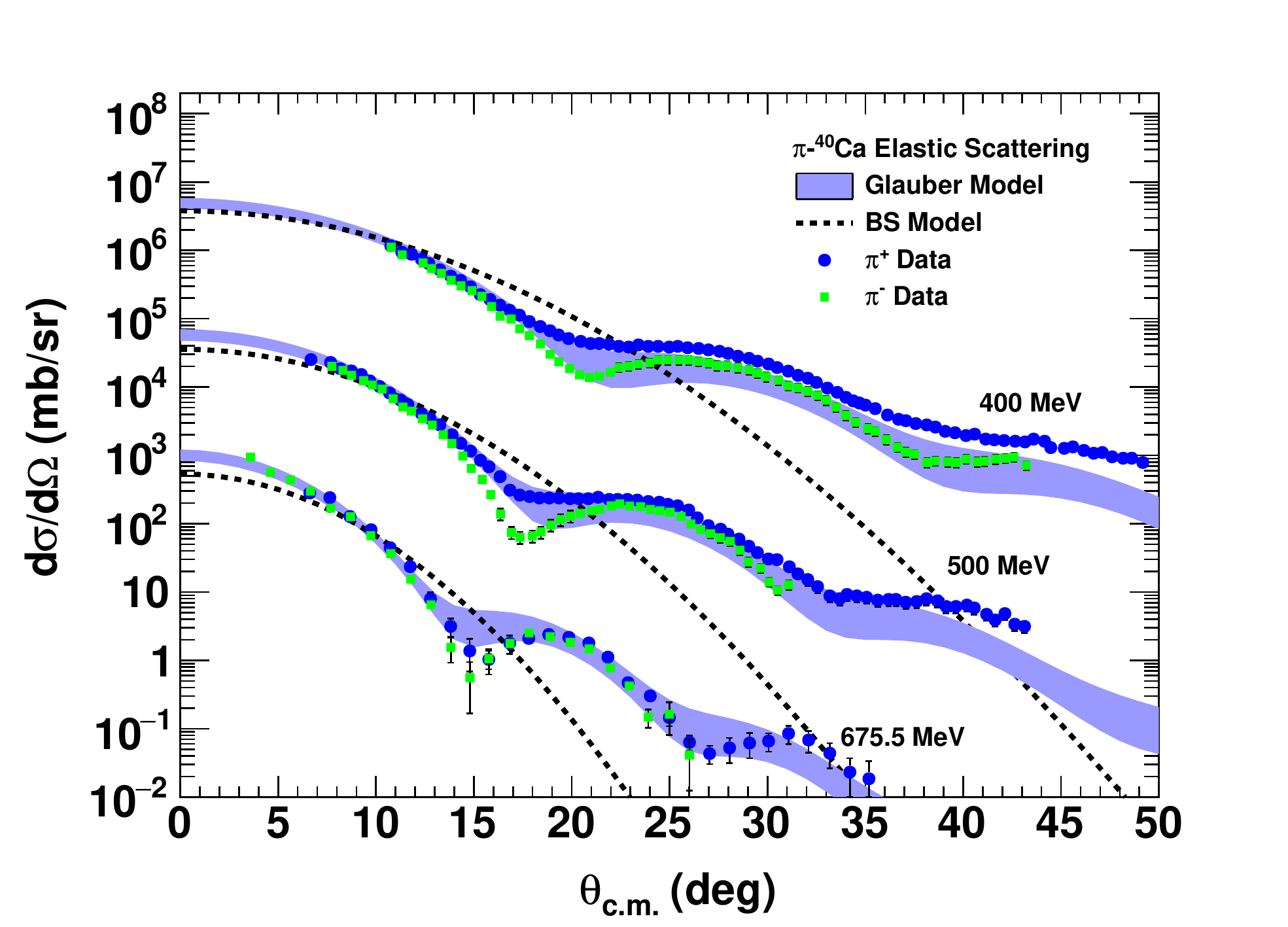}
\caption{Pion-$^{40}$Ca elastic scattering differential cross section as function of
pion scattering angle at three pion kinetic energies of 400 MeV, 500 MeV and
675.5 MeV scaled respectively by factors 1000, 10 and 0.1~\cite{Kahrimanis:1997if,BNL}.
The blue bands are obtained using the Glauber model for $\alpha_{\pi N}$ = 0.9-1.4 and
dashed lines are the calculations of the Berger-Sehgal model~\cite{Berger:2008xs}.}
\label{ElasticPionCalcium}
\end{center}
\end{figure*}

\begin{figure*}
\begin{center}
\includegraphics[width=0.65\linewidth]{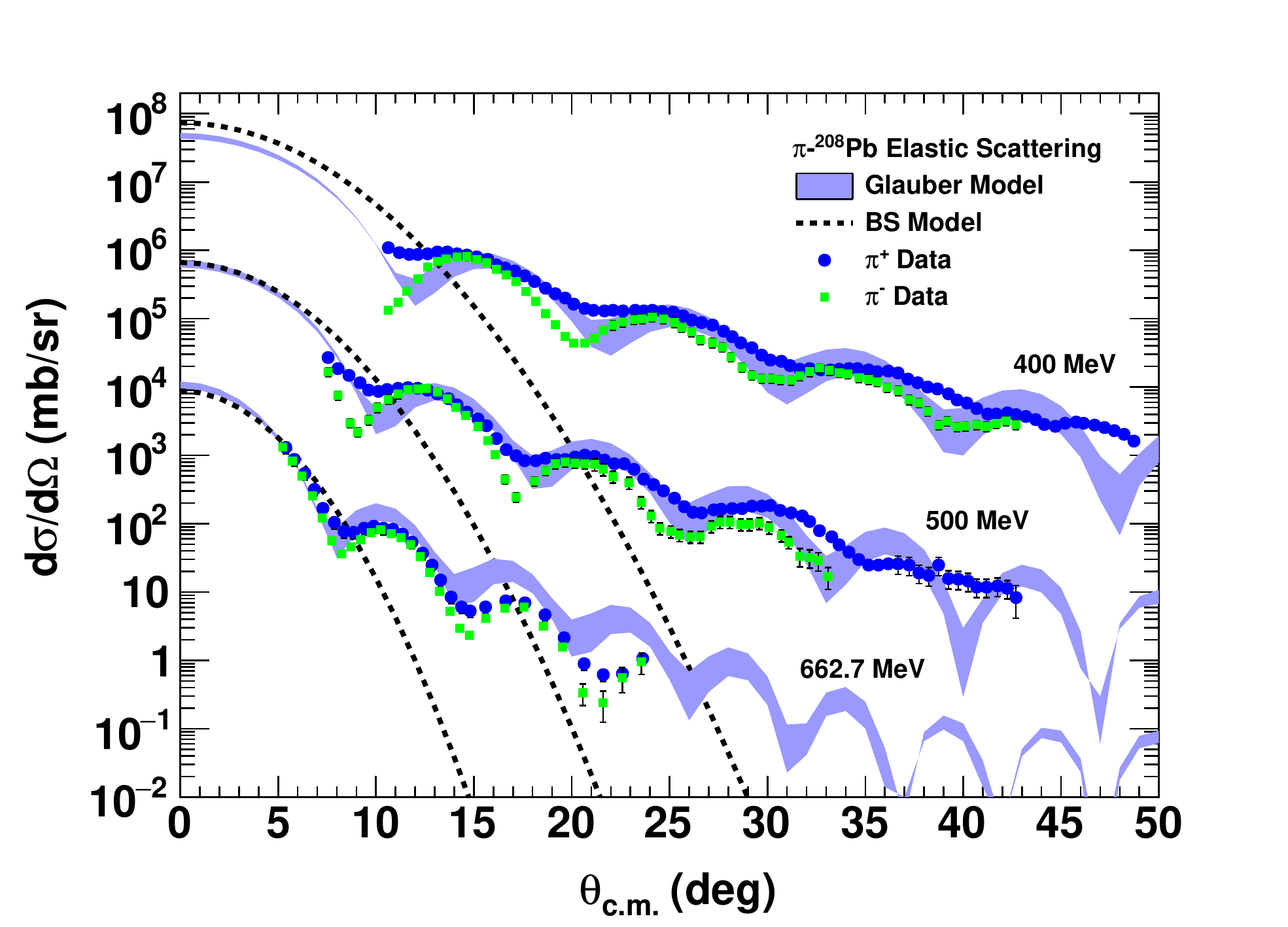}
\caption{Pion-$^{208}$Pb elastic scattering differential cross section as function of
pion scattering angle at three pion kinetic energies of 400 MeV, 500 MeV and
662.7 MeV scaled respectively by factors 1000, 10 and 0.1~\cite{Kahrimanis:1997if,BNL}.
The blue bands are obtained using the Glauber model for $\alpha_{\pi N}$ = 0.9-1.4 and
dashed lines are the calculations of the Berger-Sehgal model~\cite{Berger:2008xs}.}
\label{ElasticPionLead}
\end{center}
\end{figure*}

The values of $\alpha_{\pi N}$ obtained with the elementary scattering amplitudes
may change when one applies the formalism to nuclear targets and are typically
around one \cite{Mehndiratta:2017gry}. In view of this, we use the pion nucleus
elastic scattering data available for few incident energies of pions on a broad range of
nuclear targets to obtain $\alpha_{\pi N}$.

  Figure\,\ref{ElasticPionCarbon} shows pion-$^{12}$C elastic scattering differential
cross section as function of pion scattering angle at three pion kinetic
energies of 400 MeV, 500 MeV and 675.5 MeV scaled respectively by factors 1000, 10
and 0.1~\cite{Kahrimanis:1997if,BNL}.
The blue bands are obtained using the Glauber model for $\alpha_{\pi N}$ = 0.9-1.4 and
dashed lines are the calculations of the Berger-Sehgal model~\cite{Berger:2008xs}.

  Figure\,\ref{ElasticPionCalcium} shows pion-$^{40}$Ca elastic scattering differential
cross section as function of pion scattering angle at three pion kinetic
energies of 400 MeV, 500 MeV and 675.5 MeV scaled respectively by factors 1000, 10
and 0.1~\cite{Kahrimanis:1997if,BNL}.
The blue bands are obtained using the Glauber model for $\alpha_{\pi N}$ = 0.9-1.4 and
dashed lines are the calculations of the Berger-Sehgal model~\cite{Berger:2008xs}.

   Figure\,\ref{ElasticPionLead} shows pion-$^{208}$Pb elastic scattering differential
cross section as function of pion scattering angle at three pion kinetic
energies of 400 MeV, 500 MeV and 662.7 MeV scaled respectively by factors 1000, 10
and 0.1~\cite{Kahrimanis:1997if,BNL}.
The blue bands are obtained using the Glauber model for $\alpha_{\pi N}$ = 0.9-1.4 and
dashed lines are the calculations of the Berger-Sehgal model~\cite{Berger:2008xs}.

 It is shown that the values of $\alpha_{\pi N}$ in the range 0.9 to 1.4 give a very good
description of all the available data.  We also compare the calculations with the
Berger-Sehgal model which gives reasonable description at lower scattering angles only. 
These values of $\alpha_{\pi N}$ are then used to calculate the band
of pion production cross section in neutrino nucleus interactions.

The nuclear density function for lighter nuclei with nuclear mass upto
$^{16}$O are taken as the harmonic oscillator type as given by 
\begin{eqnarray}
\rho(r) &=& \rho_{0} \Big(1 + \alpha \, \frac{r^{2}}{a^{2}}\Big) 
\exp\Big(-\frac{r^{2}}{a^{2}}\Big), \quad
\rho_{0} = \frac{1}{(1 + 1.5 \alpha)(\sqrt{\pi}\,a)^{3}}~.
\label{carbonnucleardensity1}
\end{eqnarray}
where $a$ and $\alpha$ are given in Table\,\ref{table1}.
For light composite material $^{13.8}$CH, we use harmonic oscillator type density
with the value of $a$ = 1.729 fm (Same as that for $^{14}$N) and
$\alpha$= 1.024 calculated using formula given in Ref.~\cite{DeJager:1974liz}.
 
For nuclei heavier than $^{16}$O, the nuclear density function is taken as
two-parameter Fermi (2pF) type function as given by
\begin{eqnarray}
\rho(r) &=& \frac{\rho_{0}}{1+{\rm exp}(\frac{r-c}{d})},\quad
\rho_{0} = \frac{3}{4\pi c^{3}(1+\frac{\pi^{2}d^{2}}{c^{2}})}  ~.
\label{carbonnucleardensity2}
\end{eqnarray}
Here $d$ is the diffuseness and $c$ is the half value radius in terms of rms
radius $R_{\rm rms}$ as $c = \sqrt{(5R_{\rm rms}^{2}-7\pi^{2}d^{2})/3}$ using
the given average mass number $A$. For the composite targets with $A_{\rm eff}>16$,
the $R_{\rm rms}$ is calculated by
$R_{\rm rms} = 0.891A^{1/3}(1 + 1.565A^{-2/3} - 1.043A^{-4/3})$~\cite{friedrich}
with $d=0.537$ fm.
Table\,\ref{table1} shows the nuclear density function parameters for
Harmonic oscillator~\cite{DeJager:1974liz} and 2pF~\cite{DeJager:1987liz} for
various nuclei used in the present work.

\begin{table}[ht]
  \caption{Nuclear density function parameters for
    Harmonic oscillator~\cite{DeJager:1974liz} and
    2pF~\cite{DeJager:1987liz} density functions for various nuclei.}
\begin{center}
\begin{tabular}{|c|c|c|c|c|c|} 
\hline
\multicolumn{6}{|c|}{\bf Nuclear density parameters}\\
 \hline
 \multicolumn{3}{|c|}{\bf Harmonic oscillator}&\multicolumn{3}{|c|}{\bf 2pF} \\
\hline
Nucleus & $a$ (fm)& $\alpha$ &Nucleus & $c$ (fm) &  $d$ (fm)\\ 
\hline 
$^{7}$Li & 1.77  & 0.329 & $^{28}$Si & 3.14  & 0.537  \\ 
$^{12}$C & 1.687 & 1.029 & $^{40}$Ar & 3.53  & 0.542 \\
$^{16}$O & 1.805 & 1.446 & $^{56}$Fe & 4.106 & 0.519\\
 &  & & $^{207}$Pb & 6.62 & 0.546\\
 \hline
\end{tabular}
\end{center}
\label{table1}
\end{table}

\section{\bf Results and discussions}

In this section, we present the results of cross sections calculated using
Eq.\,\ref{marage1993} for both neutrino and anti-neutrino interactions with nucleus.
The differential cross section corresponding to an experiment is obtained by averaging
cross section over all energies weighted by the neutrino (anti-neutrino) energy
spectrum for the experiment.
\begin{equation}
  <\frac{d\sigma}{dQ^{2}}> = \frac{\int^{E_{\rm{max}}}_{E_{\rm{min}}}
    \frac{d\sigma}{dQ^{2}}(E)~ \phi(E) ~ dE}
       {\int^{E_{\rm{max}}}_{E_{\rm{min}}}  \phi(E) ~ dE}.
\label{averaged_dsigma_dQ2}
\end{equation}

\begin{figure*}
\includegraphics[width=0.48\linewidth]{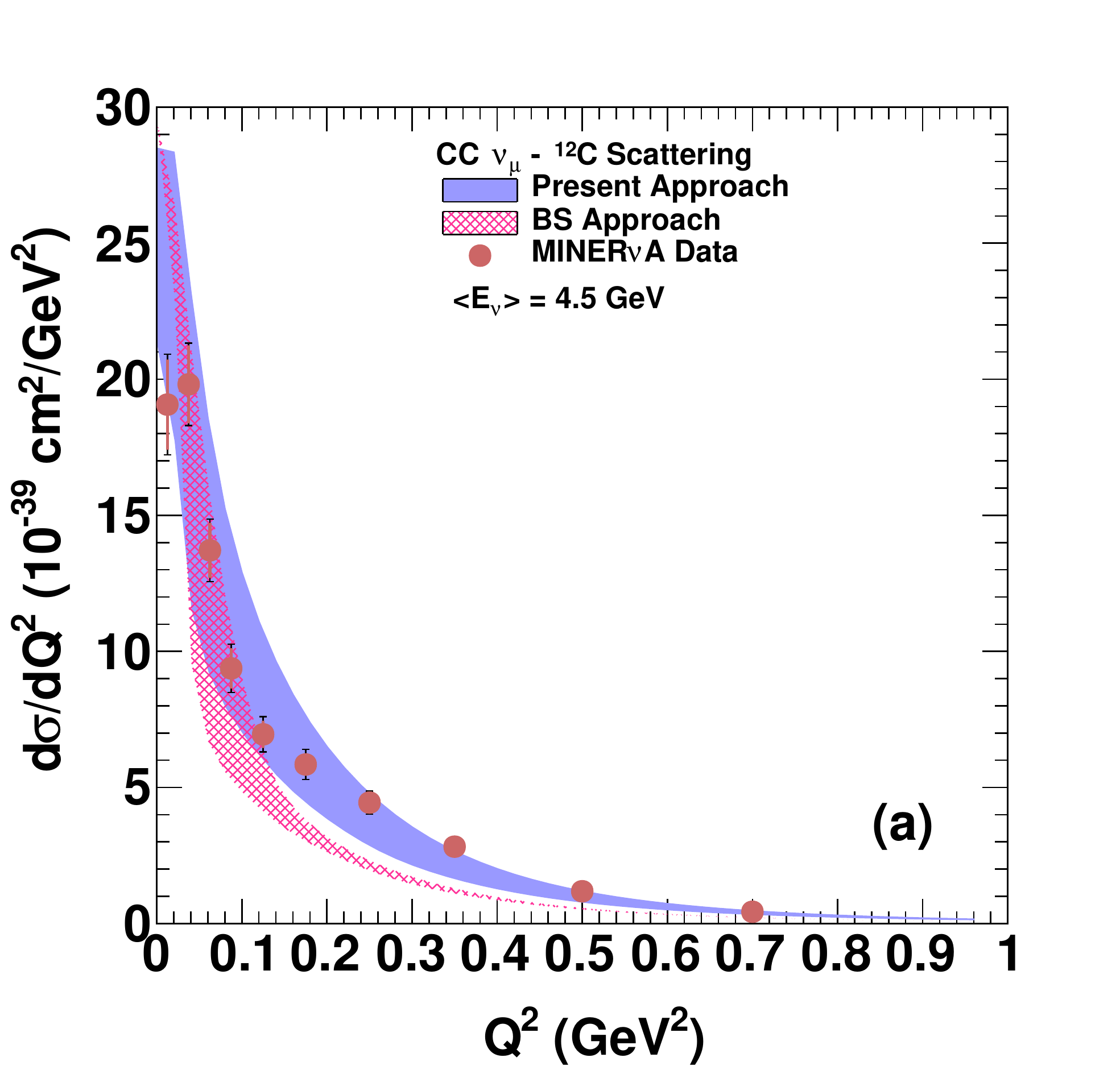}
\includegraphics[width=0.48\linewidth]{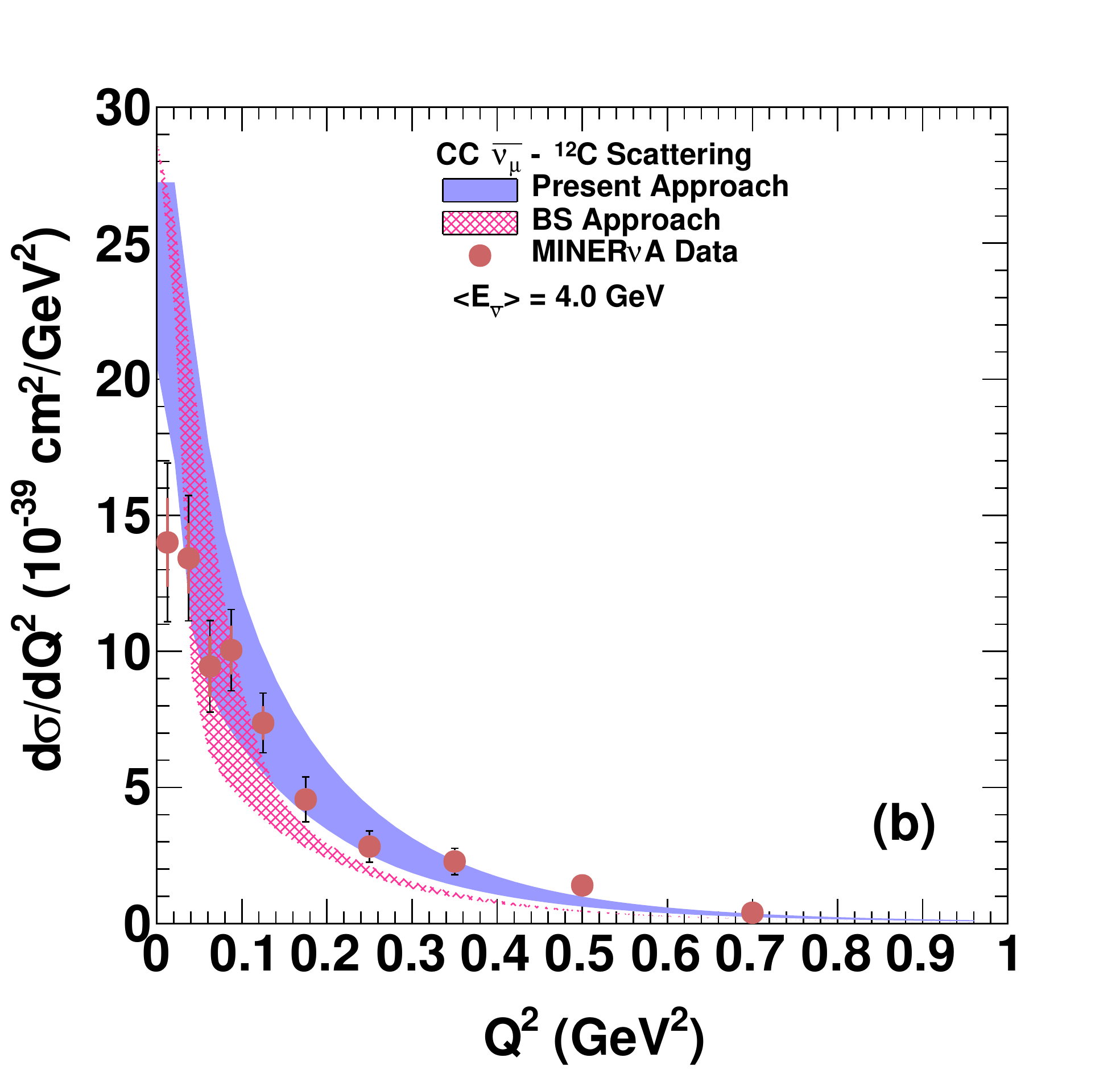}
\caption{Differential cross section $d\sigma/dQ^{2}$ (averaged over neutrino
  and anti-neutrino flux) for the charge current coherent pion production
  in $\nu_{\mu}-^{12}$C interaction as a function of the square of four-momentum transfer
  $Q^{2}$ obtained using the Glauber model based present approach and BS approach
  for (a) neutrinos with $<E_{\nu}>$ = 4.5 GeV and
  (b) anti-neutrinos with $<E_{\bar{\nu}}>$ = 4 GeV in comparison with MINER$\nu$A
  data~\cite{Mislivec:2017qfz}.} 
\label{minerva_carbon_nu_diff}
\end{figure*}

Figure\,\ref{minerva_carbon_nu_diff} (a) shows the differential cross section
$d\sigma/dQ^{2}$ (averaged over neutrino flux) for the charge current coherent
pion production in neutrino-carbon interaction as a function of the square of
four-momentum transfer $Q^{2}$ and Figure\,\ref{minerva_carbon_nu_diff}
(b) shows the same for anti-neutrino-carbon interaction
using the Glauber model based present approach and BS approach.
The band in the Glauber model corresponds to the maximum difference due to
variation of both  $\alpha_{\pi N}$ (in range 0.9-1.4) and $\xi$ (in range 1-2)
while for BS model the band includes variation of $\xi$ only.
The calculations correspond to the average energy 
4.5 GeV for neutrinos and 4 GeV for anti-neutrinos for the case 
of MINER$\nu$A experiment~\cite{Mislivec:2017qfz}.
The Glauber approach gives a better agreement (as compared to BS)
with the data especially at high $Q^2$.

\begin{figure*}
\includegraphics[width=0.48\linewidth]{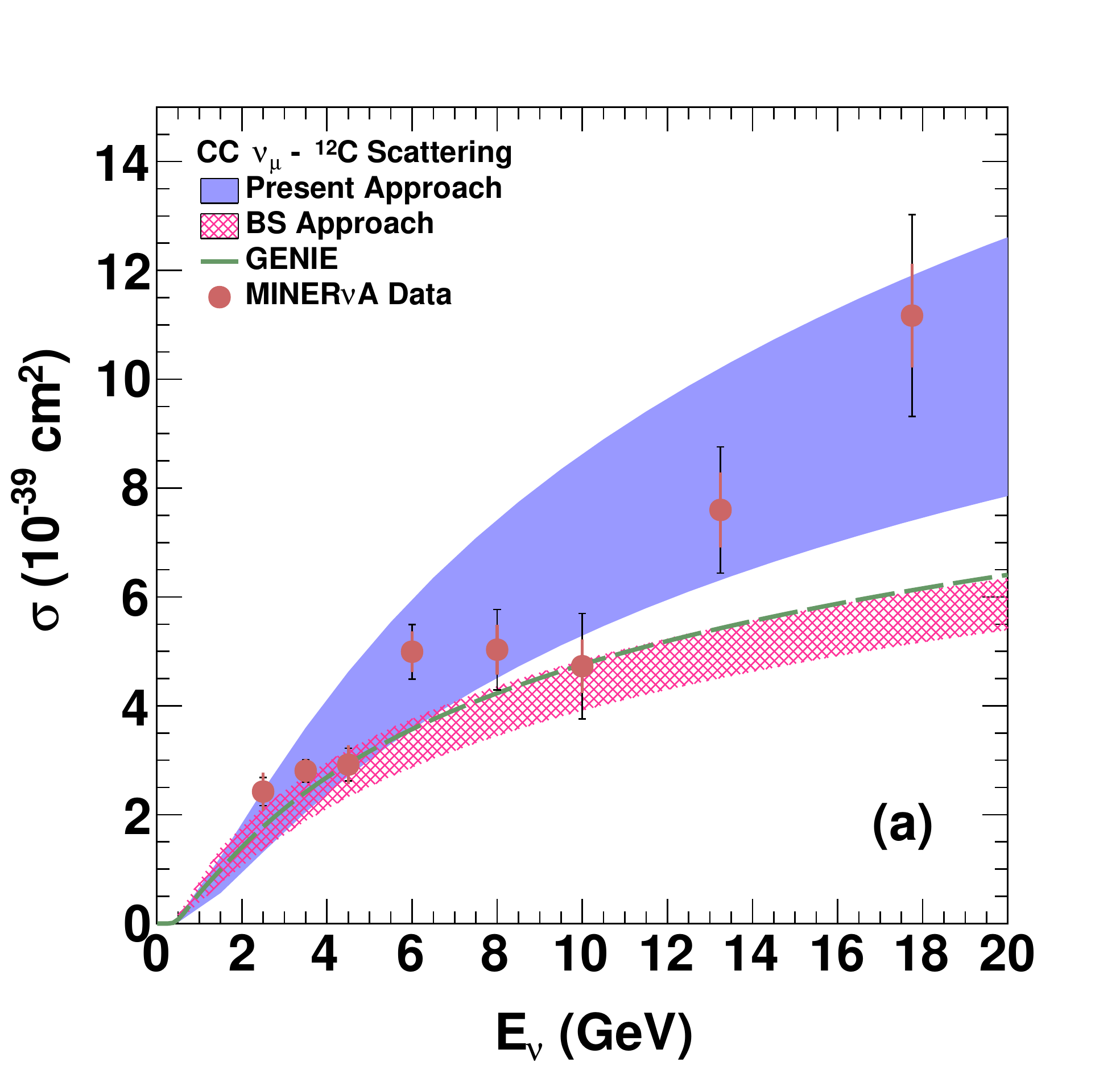}
\includegraphics[width=0.48\linewidth]{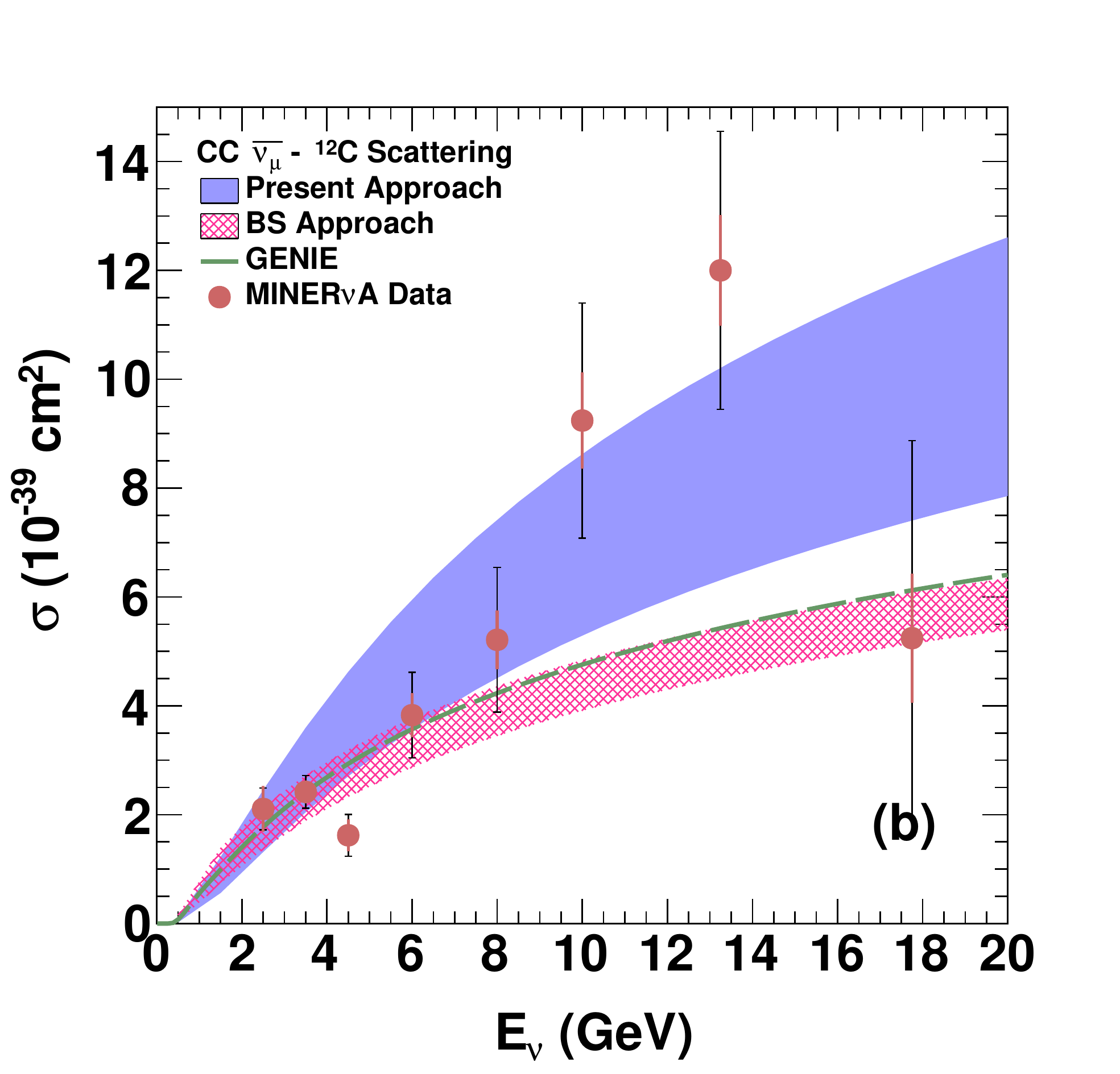}
\caption{Total cross section $\sigma$ for the charge current coherent pion
  production in $\nu_{\mu}-^{12}$C interaction as a function of neutrino
  energy $E_{\nu}$ obtained using
 the Glauber model based present approach and BS approach
  for (a) neutrinos and
  (b) anti-neutrinos. The calculations are compared with the MINER$\nu$A
  data~\cite{Mislivec:2017qfz} and GENIE.} 
\label{minerva_nu_carbon_total}
\end{figure*}

Figure\,\ref{minerva_nu_carbon_total} (a) shows the total cross section for
the charge current coherent pion production in neutrino-carbon interaction as
a function of neutrino energy and Figure\,\ref{minerva_nu_carbon_total} (b) shows
the same for anti-neutrino obtained using the Glauber model based present approach
and BS approach. 
The band in the Glauber model corresponds to the maximum difference due to
variation of both  $\alpha_{\pi N}$ (in range 0.9-1.4) and $\xi$ (in range 1-2)
while for BS model, the band includes variation of $\xi$ only.
 The calculations are compared with the measured
data of MINER$\nu$A experiment~\cite{Mislivec:2017qfz} and GENIE calculations.
The present calculations and GENIE give a good description of the data at low energies
while at high energies, the present calculation is in better agreement with data
as compared to BS approach and GENIE.

 The GENIE package uses Rein sehgal (RS) model.
 The BS model is actually RS model with improvement in the parametrization of the
 pion nucleon data.
 The difference between BS approach and GENIE
can arise due to the differences between the treatment of pion nucleon data.
 We have shown the comparison between Berger Sehgal (BS) model and Glauber model in
Figs. \ref{ElasticPionCarbon}, \ref{ElasticPionCalcium} and \ref{ElasticPionLead}.
The pion nucleus cross section goes as input in PCAC calculations and creates the differences
amoung the neutrino nucleus cross section.
 The same parameters and input pion nucleon cross sections used for Glauber model are
used in BS calculations here and thus the difference between the two is only due
to method of the modeling.
 It looks that the BS model underestimates the pion nucleus cross section
at higher energies. The Glauber model has assumptions which work better at higher
collision energies.

\begin{figure*}
\begin{center}
\includegraphics[width=0.6\textwidth]{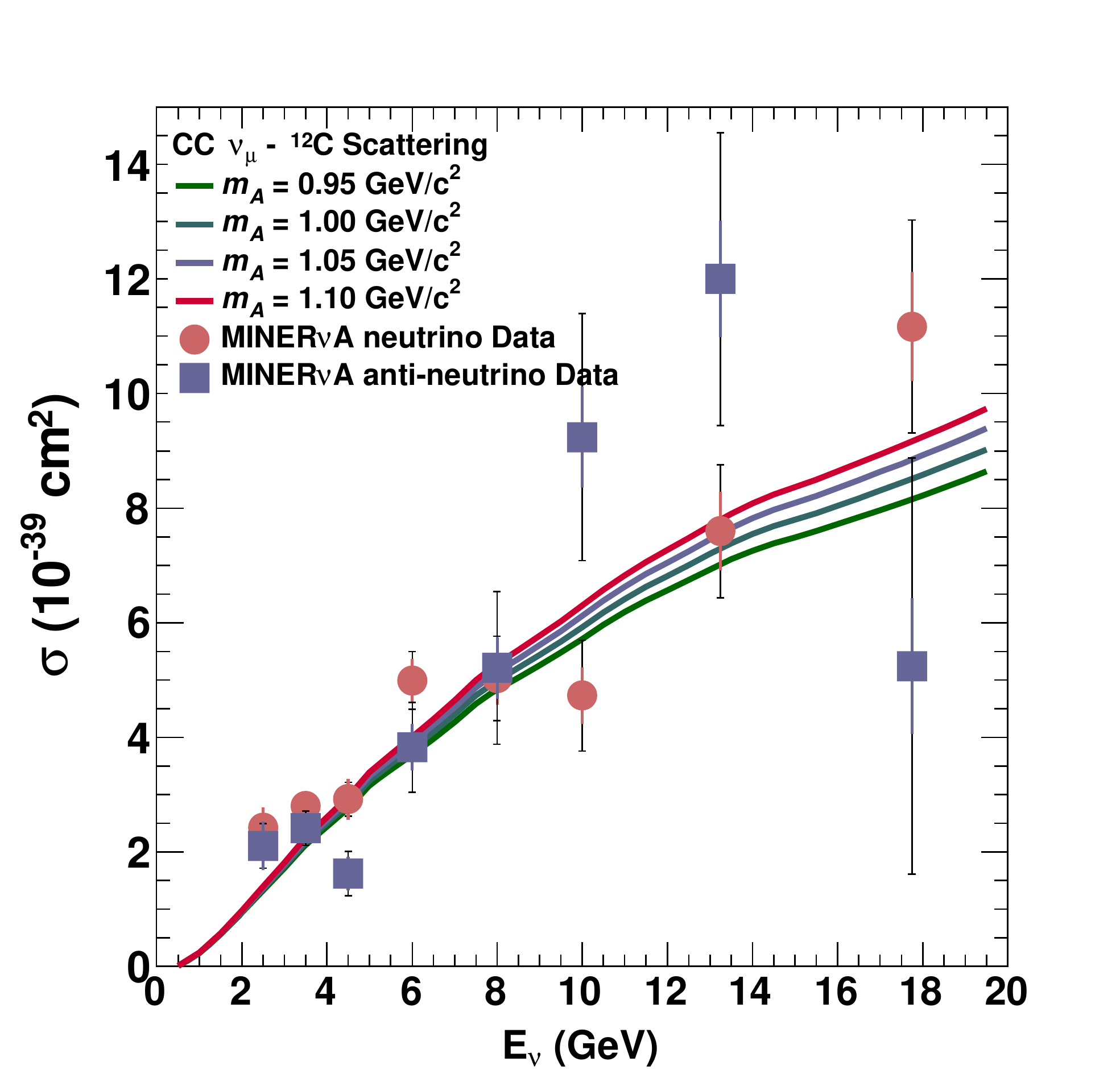}
\caption{Total cross section $\sigma$ for the charge current coherent pion
  production in $\nu_{\mu}$ or $\bar{\nu_{\mu}}-  ^{12}$C interaction as a
  function of neutrino energy $E_{\nu}$ obtained using our calculations
 ($\alpha_{\pi N}$ =1.1 and $\xi$=1) for four
  different values of axial mass parameter $m_{A}$ in comparison with
  MINER$\nu$A experimental data~\cite{Mislivec:2017qfz}.} 
\label{MA}
\end{center}
\end{figure*}
Figure\,\ref{MA} shows the total cross section for the charge current coherent
pion production neutrino (anti-neutrino)-carbon interaction as a
function of neutrino energy obtained using the Glauber model based present
approach with $\alpha_{\pi N}$ =1.1 and $\xi$=1.
The calculations are done for four different values of
axial mass parameter $m_{A}$ and are compared with 
MINER$\nu$A experimental data~\cite{Mislivec:2017qfz}.
Different values of $m_A$ make difference only at high energies where there are
large uncertainties of the measurements. We have chosen $m_A=1.05$ GeV/$c^2$ in all
our calculations.

\begin{figure*}
\includegraphics[width=0.48\linewidth]{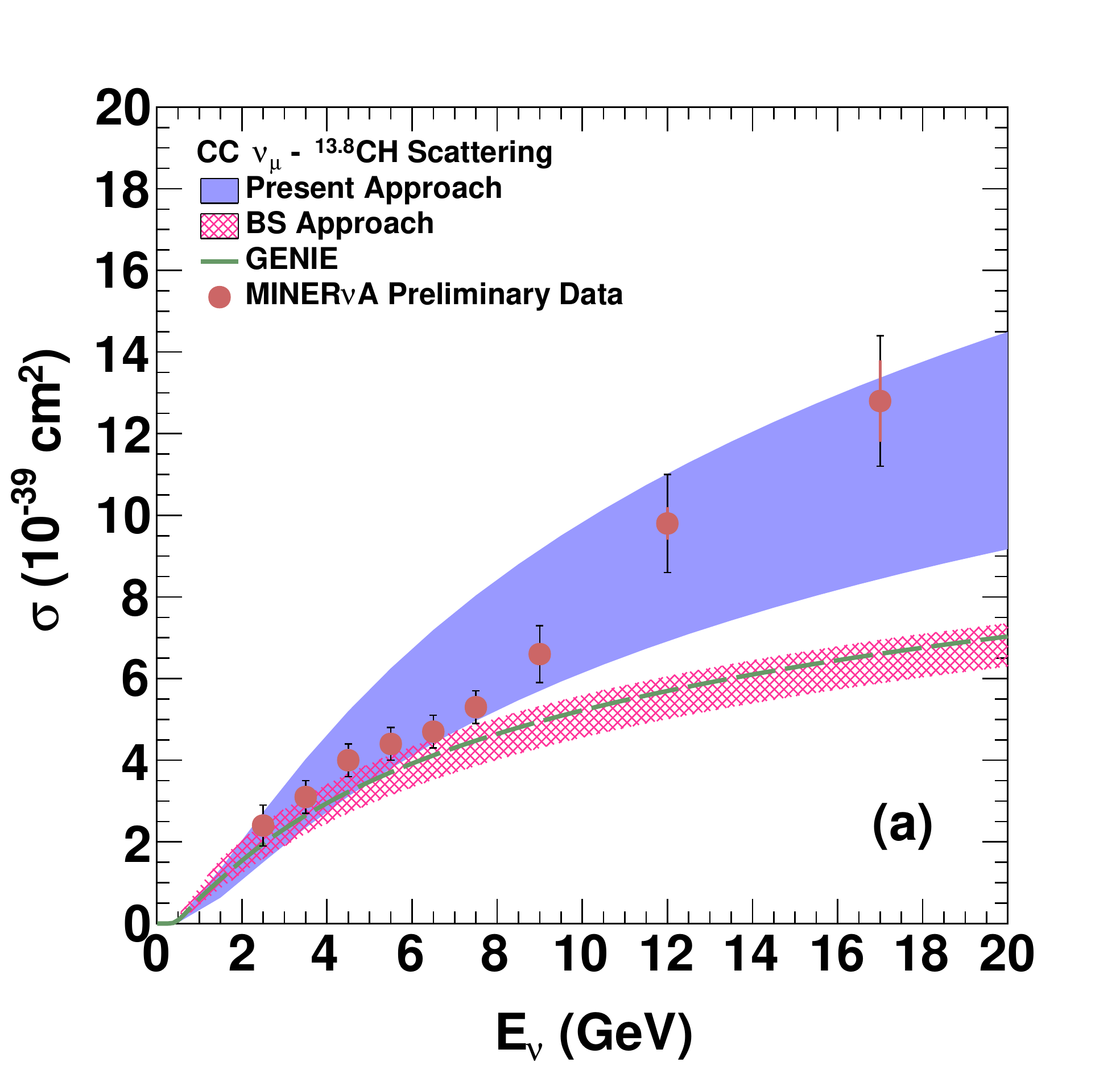}
\includegraphics[width=0.48\linewidth]{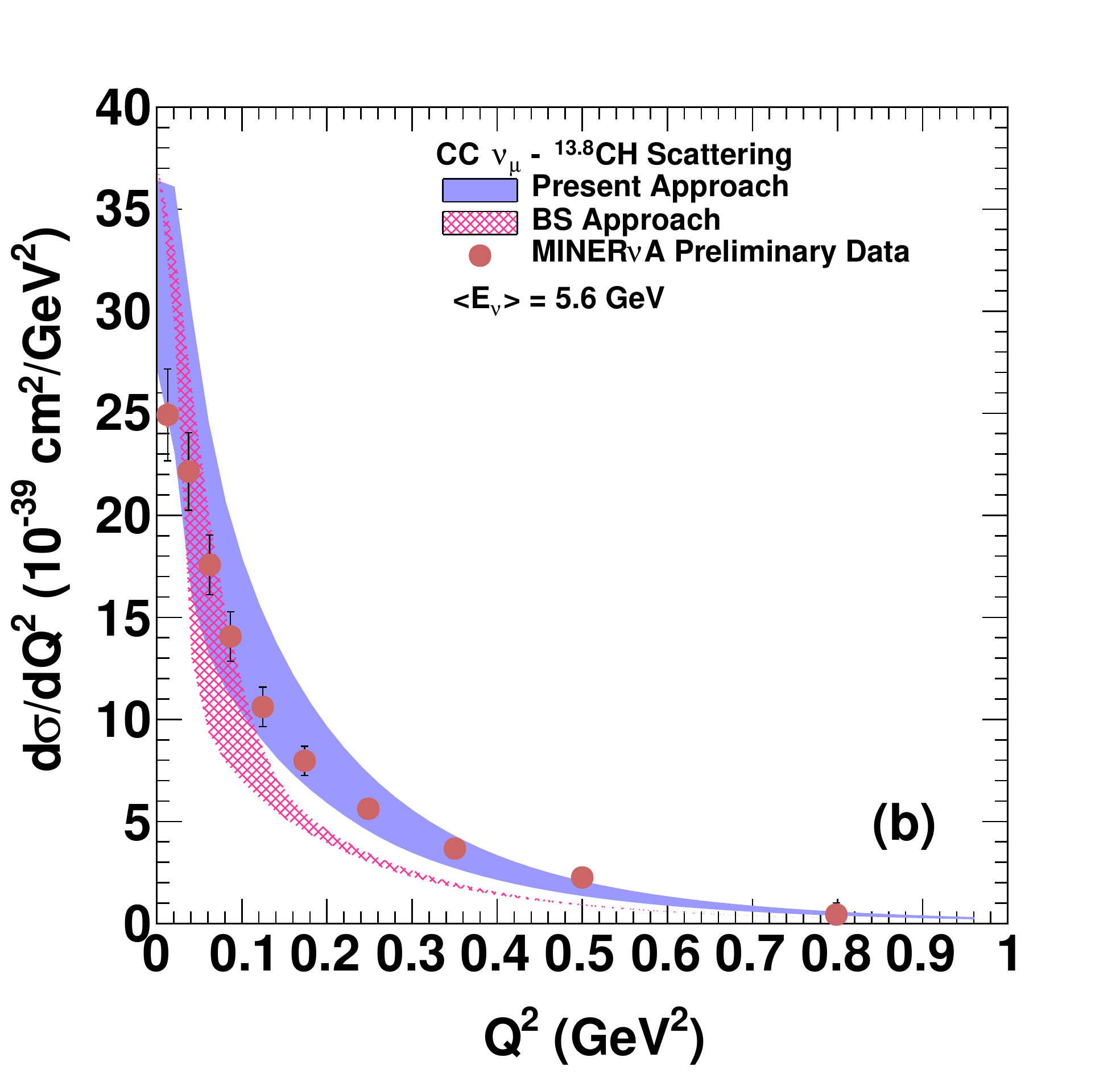}
\caption{Charge current coherent pion production in $\nu_{\mu}-^{13.8}$CH
  interaction obtained using the Glauber model based present approach
and BS approach
  (a) Total cross section $\sigma$
as a function of neutrino energy $E_{\nu}$ compared with
the preliminary MINER$\nu$A data~\cite{tminervadata} and GENIE
and (b) Differential cross section
  (averaged over neutrino flux~\cite{tminervadata}) as a function of the square of
  four-momentum transfer $Q^{2}$ compared with the 
  preliminary MINER$\nu$A data~\cite{tminervadata}.
}
\label{tminerva_nu_ch_total}
\end{figure*}

Figure\,\ref{tminerva_nu_ch_total} (a) shows the total cross section for the
charge current coherent pion production in neutrino-hydrocarbon interaction
as a function of neutrino energy obtained using
the Glauber model based present approach and BS approach
compared with the preliminary data of MINER$\nu$A
experiment~\cite{tminervadata} and GENIE calculations.
 Figure\,\ref{tminerva_nu_ch_total}
(b) shows differential cross section  (averaged over neutrino flux~\cite{tminervadata})
 as a function of the square of four-momentum transfer $Q^{2}$
 obtained using the Glauber model based present approach and BS approach
 compared with the preliminary data of MINER$\nu$A
 experiment~\cite{tminervadata}.
 The present calculations and GENIE give
good description of the total cross section data at low energies while at high energies our calculation
is in better agreement with data as compared to GENIE and BS approach.
The present calculations of the differential cross section match much better with the data over
whole $Q^2$ range as compared to BS approach.
The cross section by GENIE shown in the figure is obtained by
scaling their result on carbon with $(13.8/12)^{\frac{2}{3}}$.

\begin{figure*}
\includegraphics[width=0.48\linewidth]{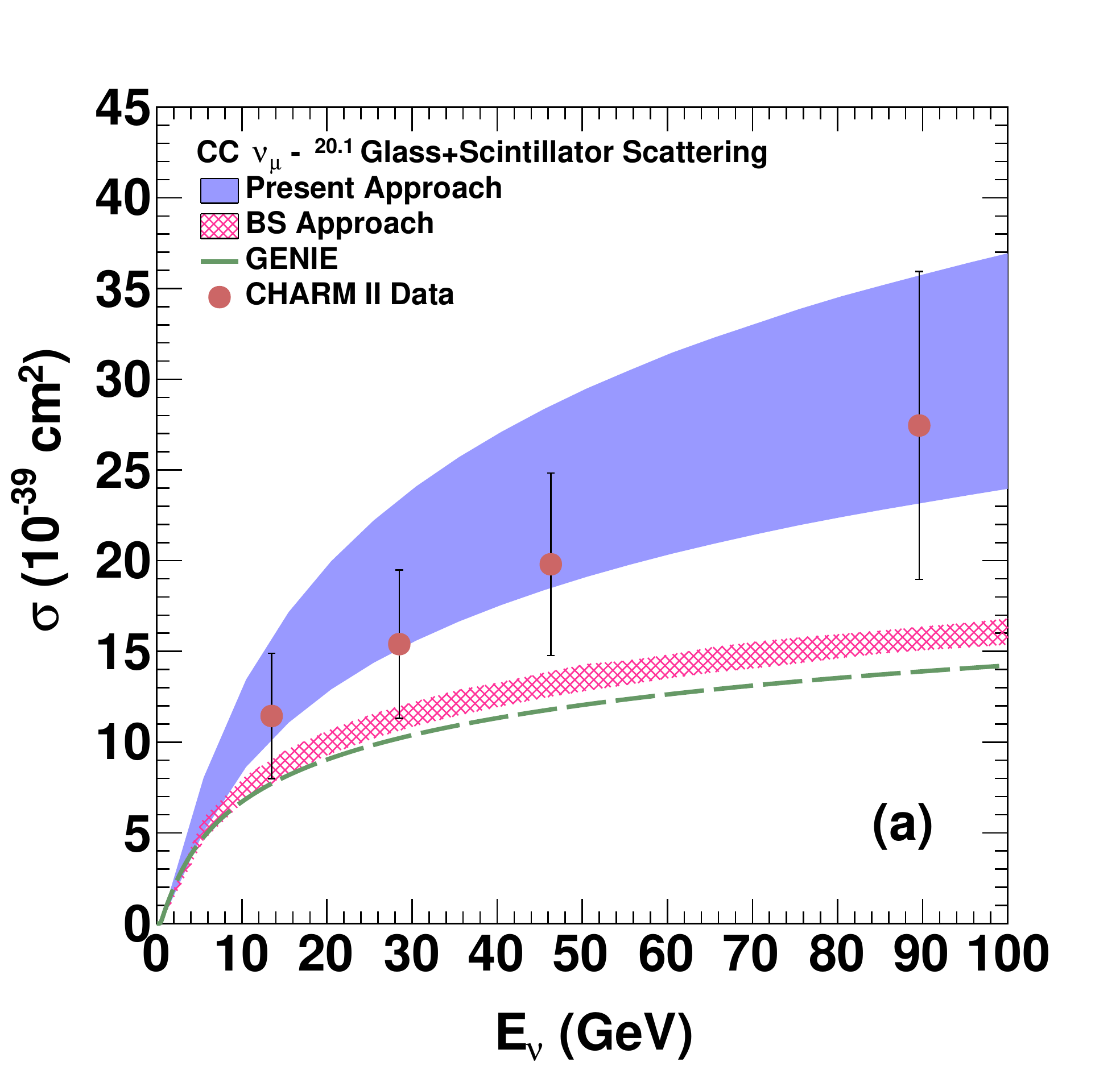}
\includegraphics[width=0.48\linewidth]{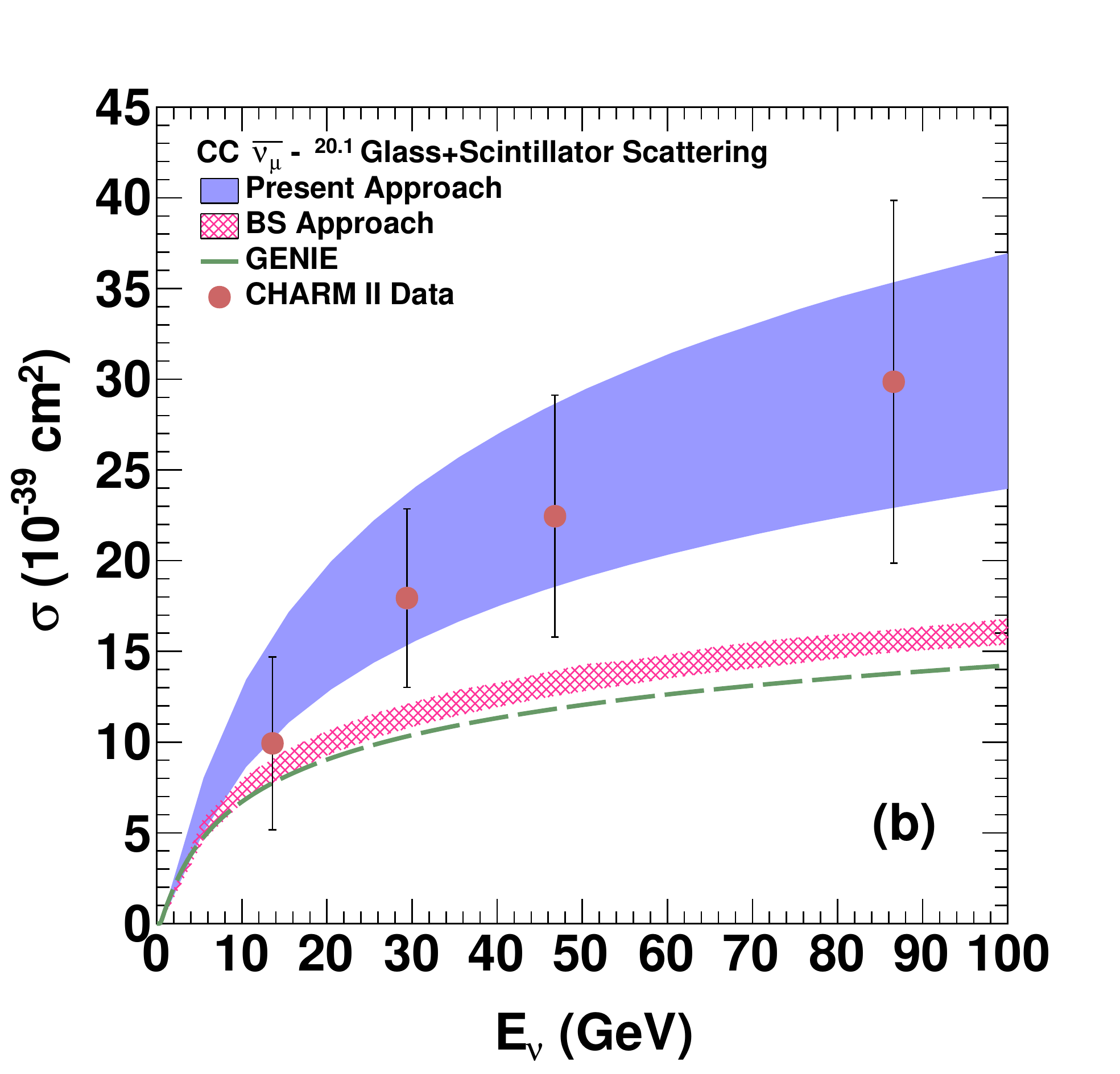}
\caption{Total cross section $\sigma$ for the charge current coherent pion
  production in neutrino-$^{20.1}$ Glass+Scintillator interaction as a
  function of neutrino energy $E_{\nu}$ obtained using
the Glauber model based present approach
and BS approach
  for (a) neutrino and (b) anti-neutrino. The calculations are compared with the
  CHARM-II data~\cite{Vilain:1993sf} and GENIE.}
\label{charmII_nu_total}
\end{figure*}

Figure\,\ref{charmII_nu_total} (a) shows the total cross section for the charge
current coherent pion production in neutrino-glass+scintillator interaction as a
function of neutrino energy obtained using
the Glauber model based present approach and BS approach.
Figure\,\ref{charmII_nu_total} (b) shows the same for anti-neutrino. The calculations
are compared with the measured data of CHARM-II experiment~\cite{Vilain:1993sf} and
GENIE. The present calculations are in better agreement with data within experimental
error as compared to GENIE and BS approach.

\begin{figure*}
\includegraphics[width=0.48\linewidth]{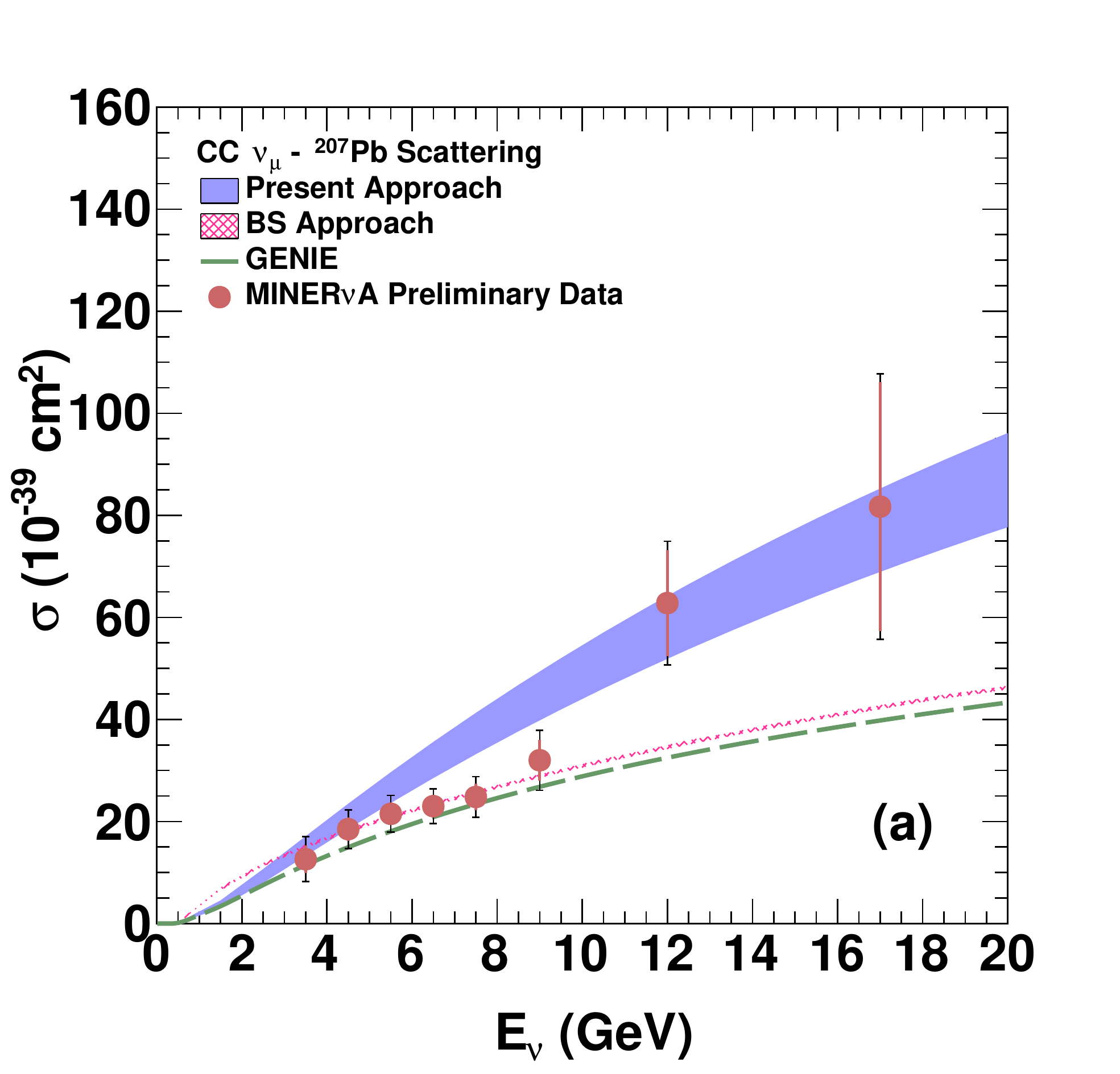}
\includegraphics[width=0.48\linewidth]{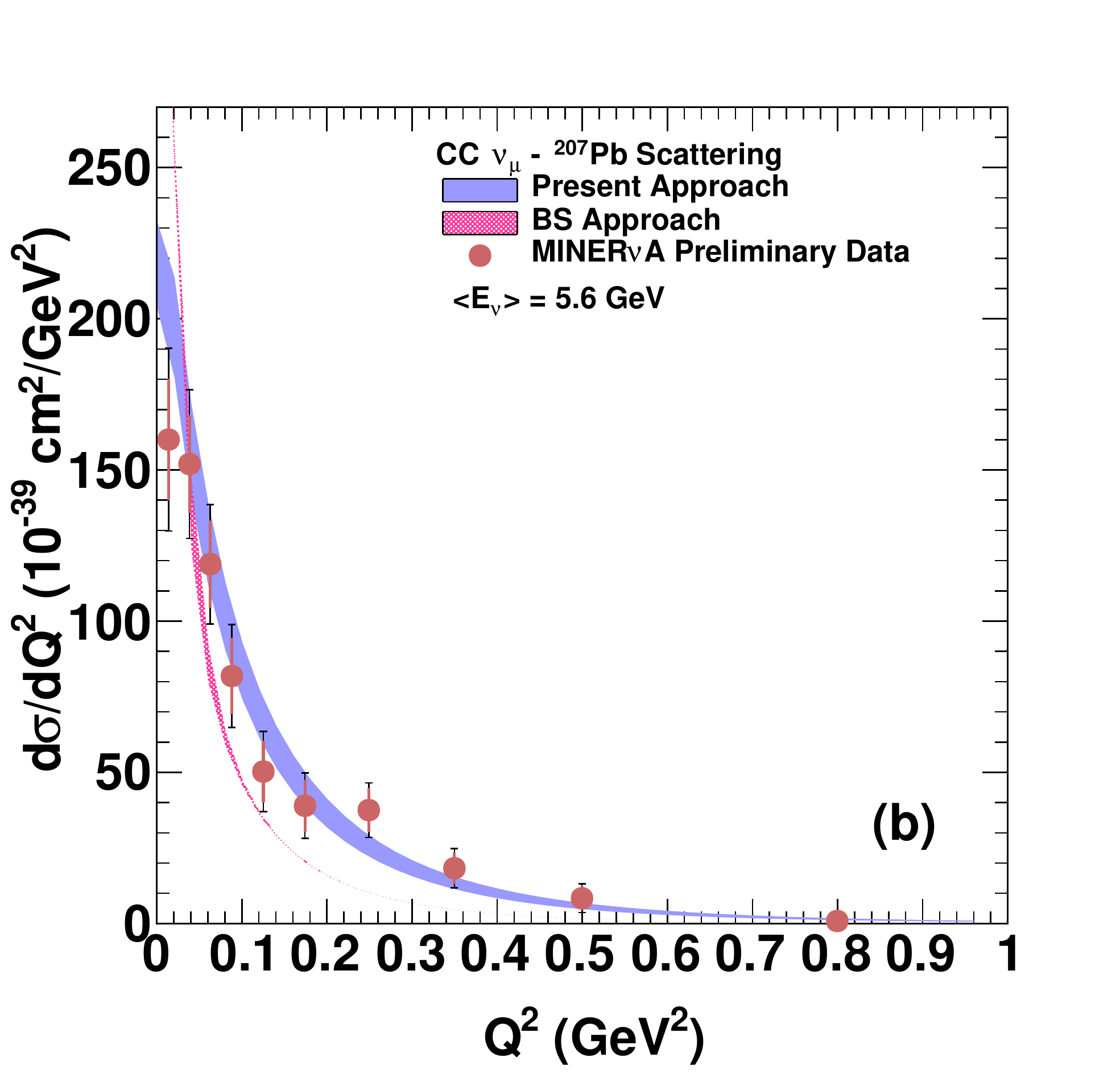}
\caption{Charge current coherent pion production in $\nu_{\mu}-^{207}$Pb
  interaction obtained using
  the Glauber model based present approach and BS approach
   compared with the
  MINER$\nu$A data~\cite{tminervadata} and GENIE.
  (a) Total cross section
  $\sigma$ as a function of neutrino energy and (b) Differential cross section
  (averaged over neutrino flux) as a function of the square of four-momentum
  transfer $Q^{2}$~\cite{tminervadata} compared with the
  MINER$\nu$A data~\cite{tminervadata}.} 
\label{tminerva_nu_pb_total}
\end{figure*}

Figure\,\ref{tminerva_nu_pb_total} (a) shows the total cross section for the
charge current coherent pion production in neutrino-lead interaction
as a function of neutrino energy obtained using
the Glauber model based present approach and BS approach
compared with the measured data of MINER$\nu$A
experiment~\cite{tminervadata} and GENIE calculations.
Figure\,\ref{tminerva_nu_pb_total}
(b) shows differential cross section  (averaged over neutrino flux~\cite{tminervadata})
as a function of the square of four-momentum transfer $Q^{2}$ compared with the measured data of MINER$\nu$A
experiment~\cite{tminervadata}. The present calculations and GENIE
give a good description of the total cross section at low energies while at high energies our
calculation is in better agreement with data as compared to GENIE and BS model.
The present calculations of the differential cross section match much better with the data over
whole $Q^2$ range as compared to BS approach.

\begin{figure*}
\includegraphics[width=0.48\linewidth]{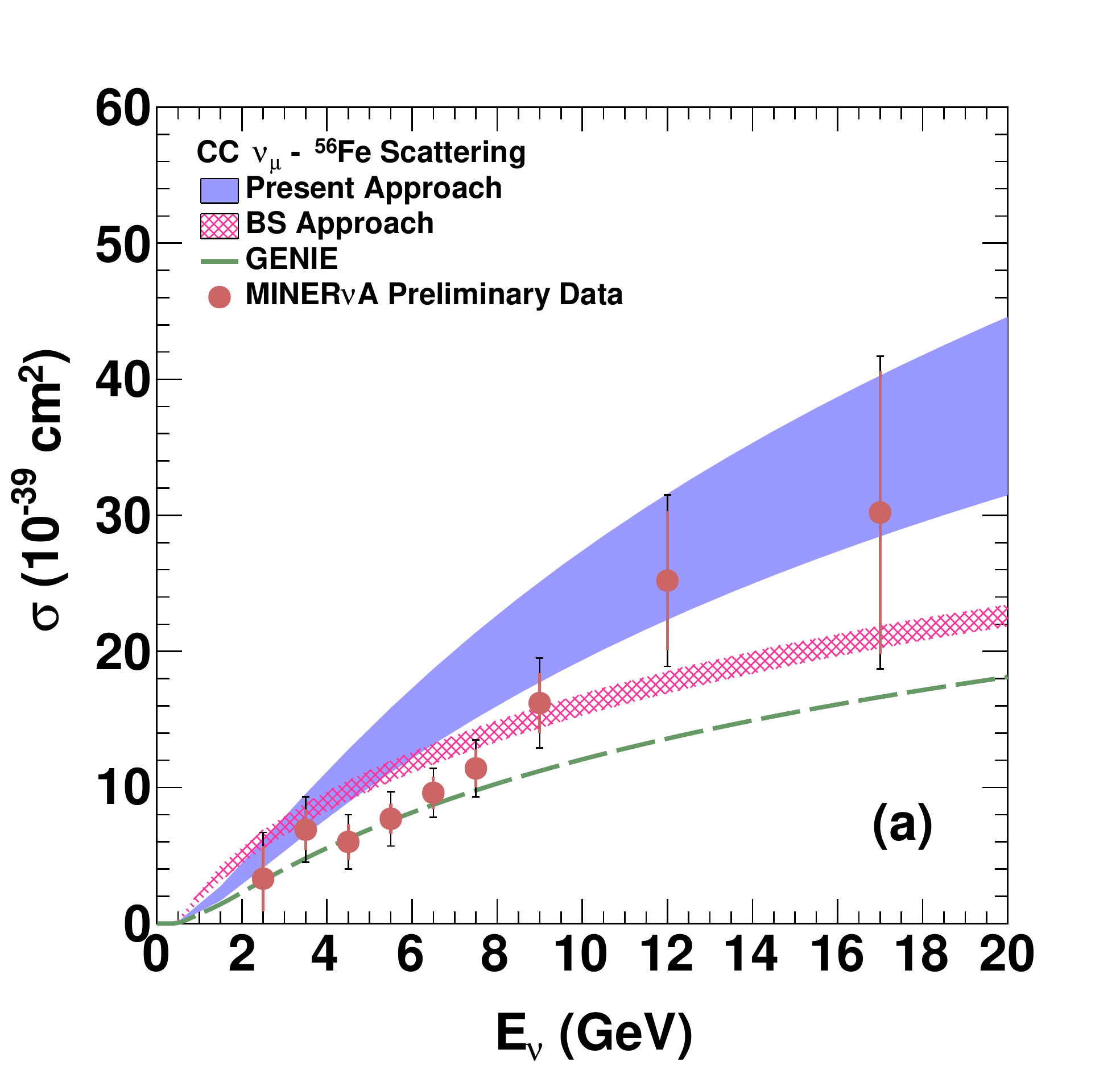}
\includegraphics[width=0.48\linewidth]{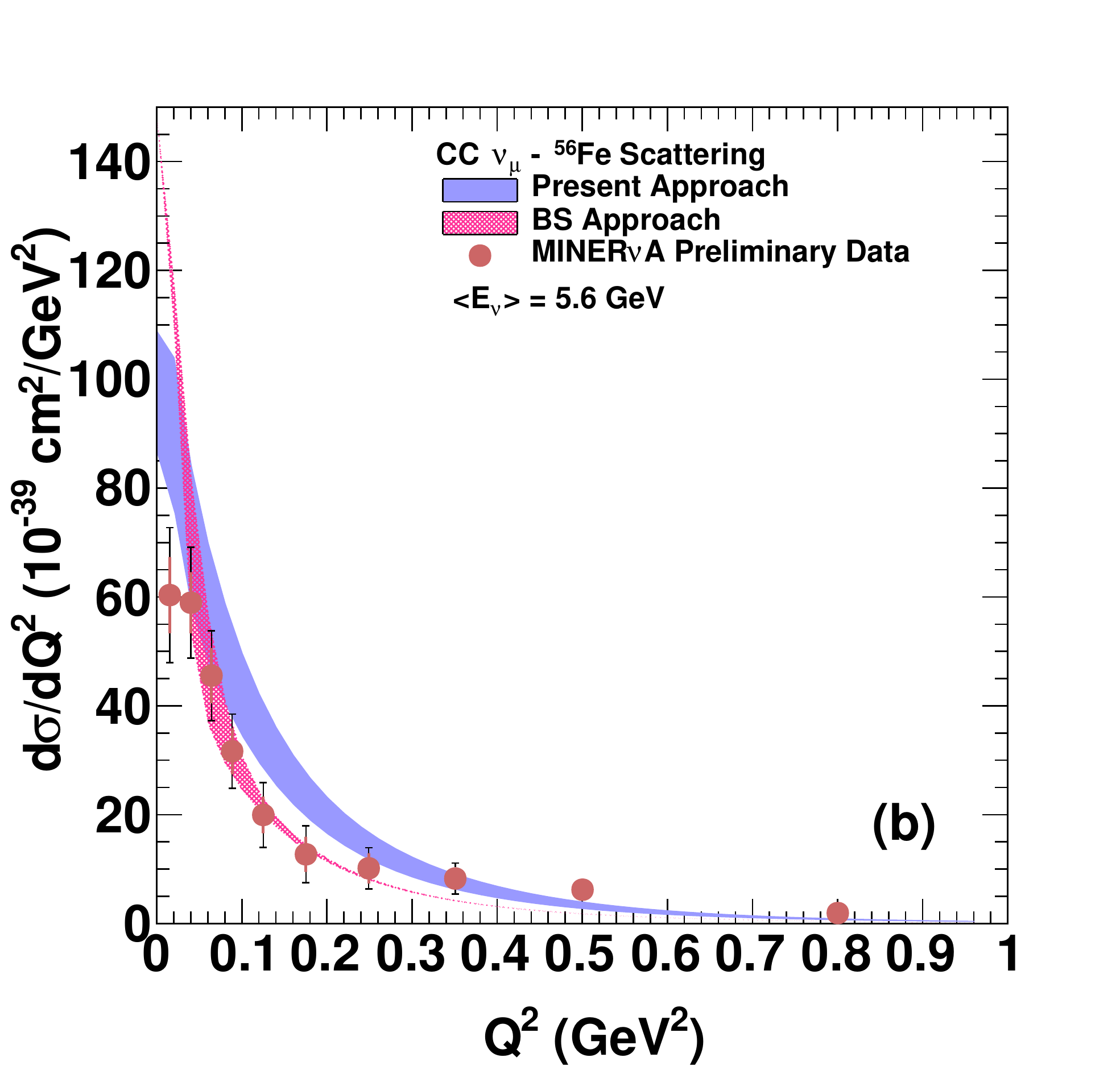}
\caption{Charge current coherent pion production in  $\nu_{\mu}-^{56}$Fe
  interaction obtained using
the Glauber model based present approach and BS approach
  (a) Total cross section
  $\sigma$ as a function of neutrino energy are compared with MINER$\nu$A
  data~\cite{tminervadata} and GENIE. (b) Differential cross section (averaged
  over neutrino flux) as a function of the square of four-momentum transfer $Q^{2}$ are
  compared with the MINER$\nu$A data~\cite{tminervadata}.} 
\label{tminerva_nu_fe_total_cc}
\end{figure*}

Figure\,\ref{tminerva_nu_fe_total_cc} (a) shows the total cross section for
the charge current coherent  pion production in neutrino-iron interaction
as a function of neutrino energy obtained using
the Glauber model based present approach and BS approach
compared with the data of MINER$\nu$A experiment~\cite{tminervadata}
and GENIE. This result is relevant for INO as well.
Figure\,\ref{tminerva_nu_fe_total_cc} (b) shows differential
cross section (averaged over neutrino flux) as a function of the square of
four-momentum transfer $Q^{2}$~\cite{tminervadata}
compared with the data of MINER$\nu$A experiment~\cite{tminervadata}.
 The GENIE and BS give a good description of the data at lower energies
while our calculations give reasonable description in the whole energy range.

\begin{figure*}
\includegraphics[width=0.48\linewidth]{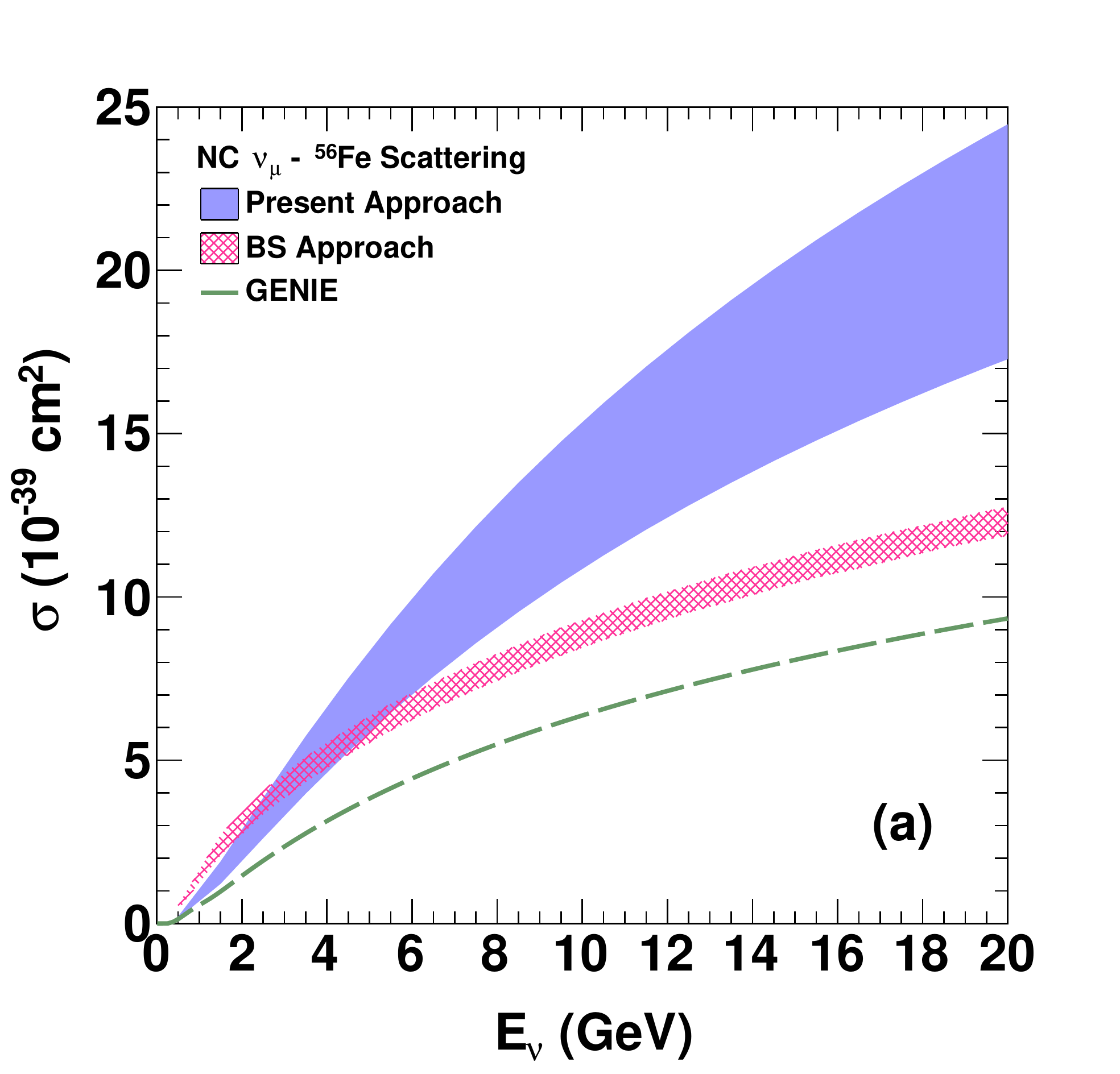}
\includegraphics[width=0.48\linewidth]{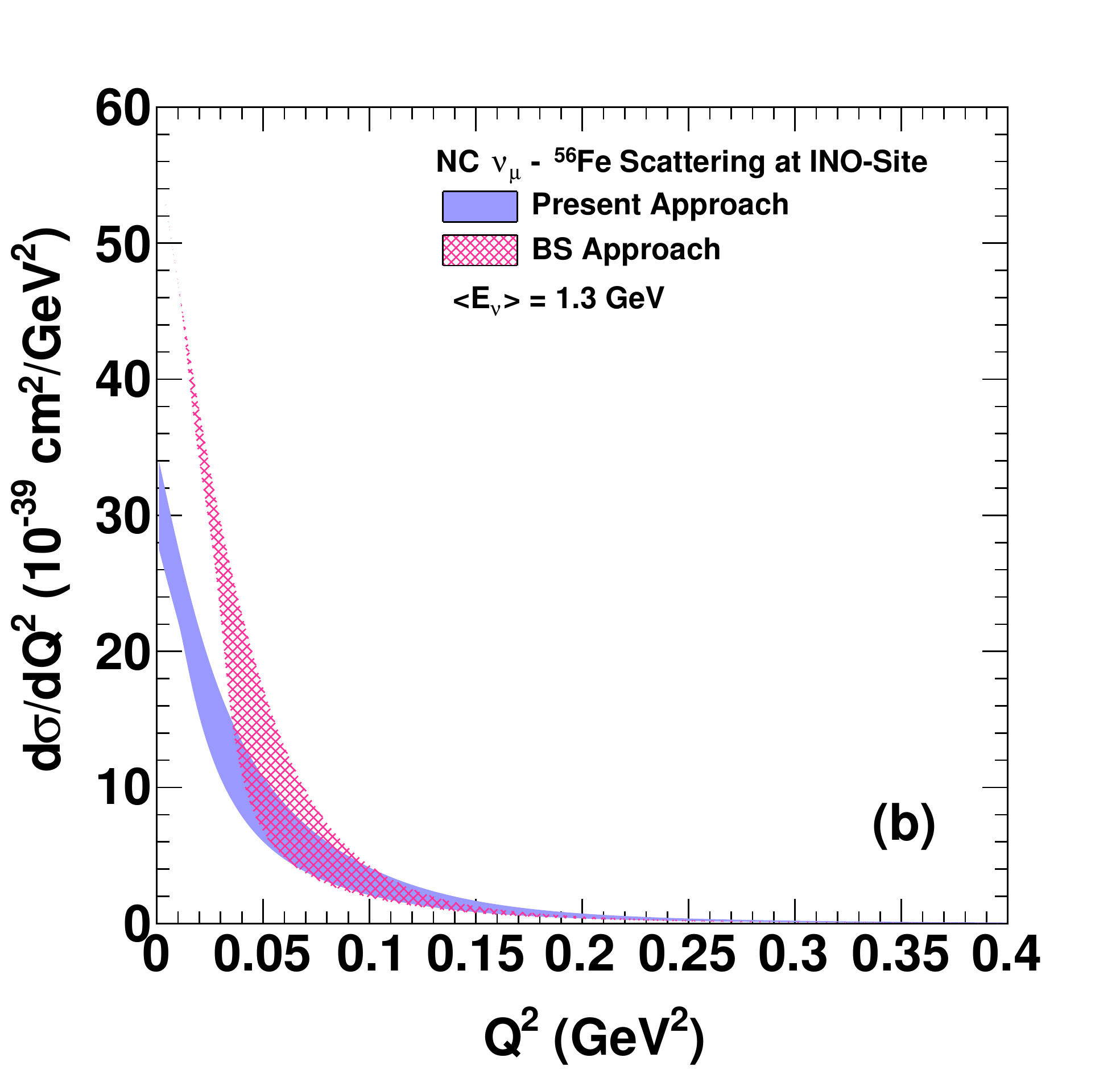}
\caption{Neutral current coherent pion production in $\nu_{\mu}-^{56}$Fe
  interaction obtained using
the Glauber model based present approach and BS approach
(a) Total cross section $\sigma$
  as a function of neutrino energy compared with GENIE and (b) Differential cross section (averaged
  over neutrino flux at INO-site ~\cite{Ahmed:2015jtv} for solar minimum~\cite{hondaflux})
  as a function of
  the square of four-momentum transfer $Q^{2}$.
}
\label{ino_nu_fe_total_nc}
\end{figure*}

\begin{figure*}
\begin{center}
\includegraphics[width=0.55\textwidth]{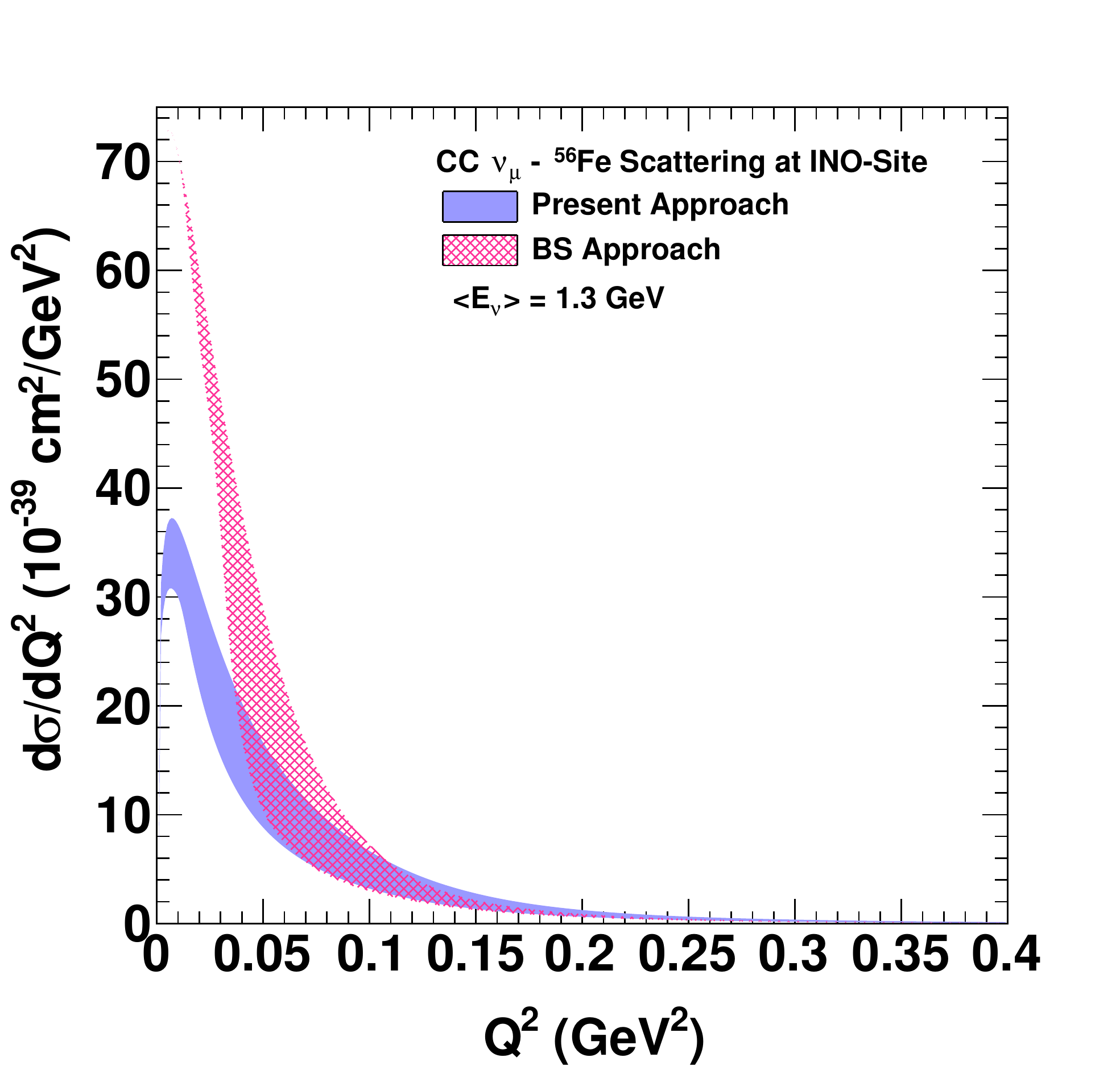}
\caption{Charge current coherent pion production in $\nu_{\mu}-^{56}$Fe
  interaction obtained using
  the Glauber model based present approach and BS approach
  for differential cross section
  (averaged over neutrino flux at INO-site for solar minimum~\cite{hondaflux}) as a
  function of the square of four-momentum transfer $Q^{2}$.
}
\label{ino_nu_fe}
\end{center}
\end{figure*}

\vspace{2cm}

Figure\,\ref{ino_nu_fe_total_nc} (a) shows the total cross section prediction for the
neutral current coherent pion production in neutrino-iron interaction relevant for INO as
a function of neutrino energy obtained using
the Glauber model based present approach and BS approach.
The calculations are compared with GENIE. 
Figure\,\ref{ino_nu_fe_total_nc} (b) shows differential
cross section (averaged over neutrino flux at INO-site~\cite{Ahmed:2015jtv} for solar
minimum from
1 GeV to 20 GeV energy range~\cite{hondaflux}) as a function of the square of four-momentum
transfer $Q^{2}$  calculated using Glauber based
approach and BS approach. 

Figure\,\ref{ino_nu_fe} shows differential cross section for charge current
coherent pion production in neutrino-iron interaction as a function of the square of
four-momentum transfer $Q^{2}$ (averaged over neutrino flux at INO-site for solar
minimum from 1 GeV to 20 GeV energy range~\cite{hondaflux}) calculated using Glauber based
approach and BS approach.

\begin{figure*}
\includegraphics[width=0.48\linewidth]{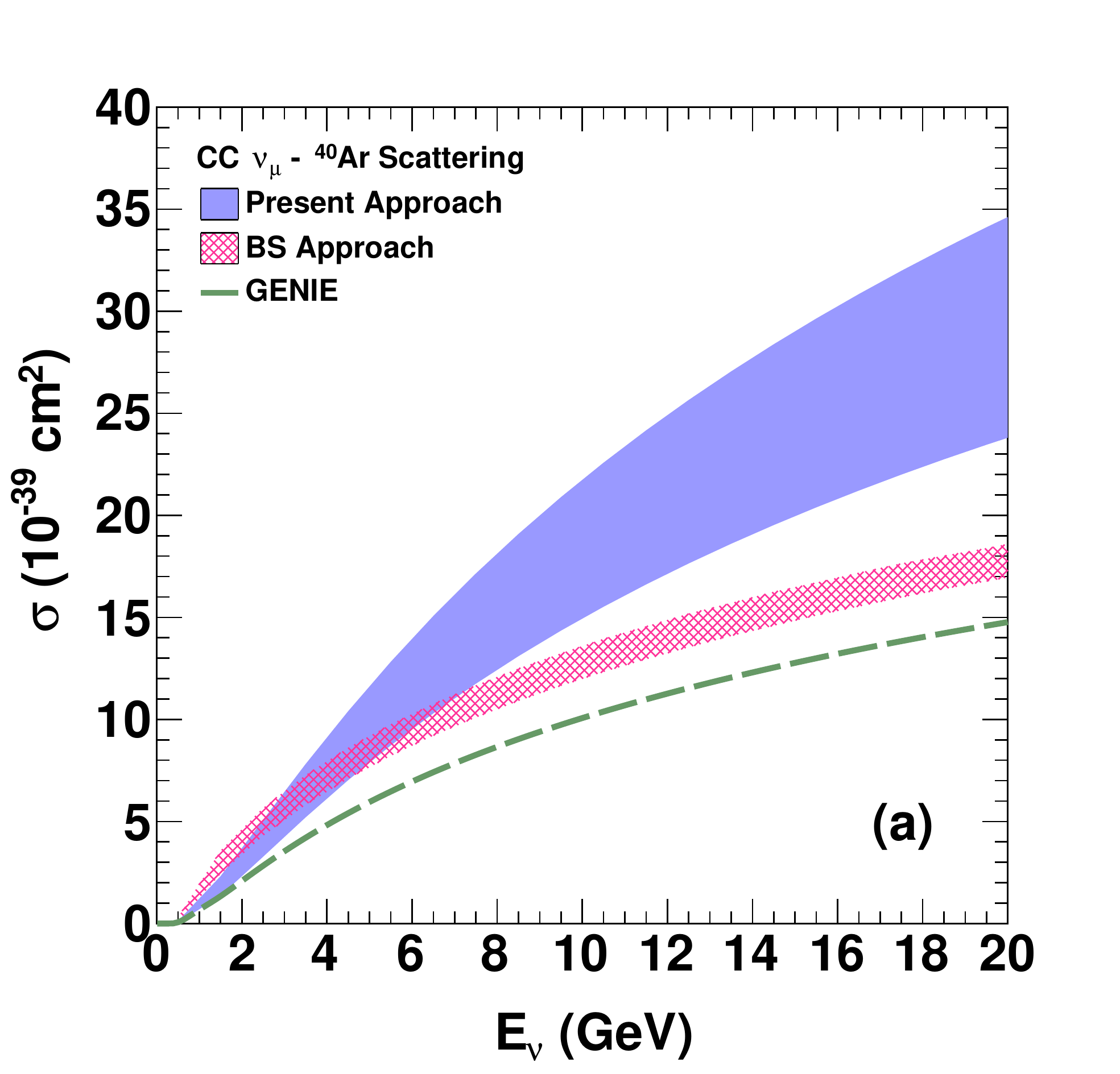}
\includegraphics[width=0.48\linewidth]{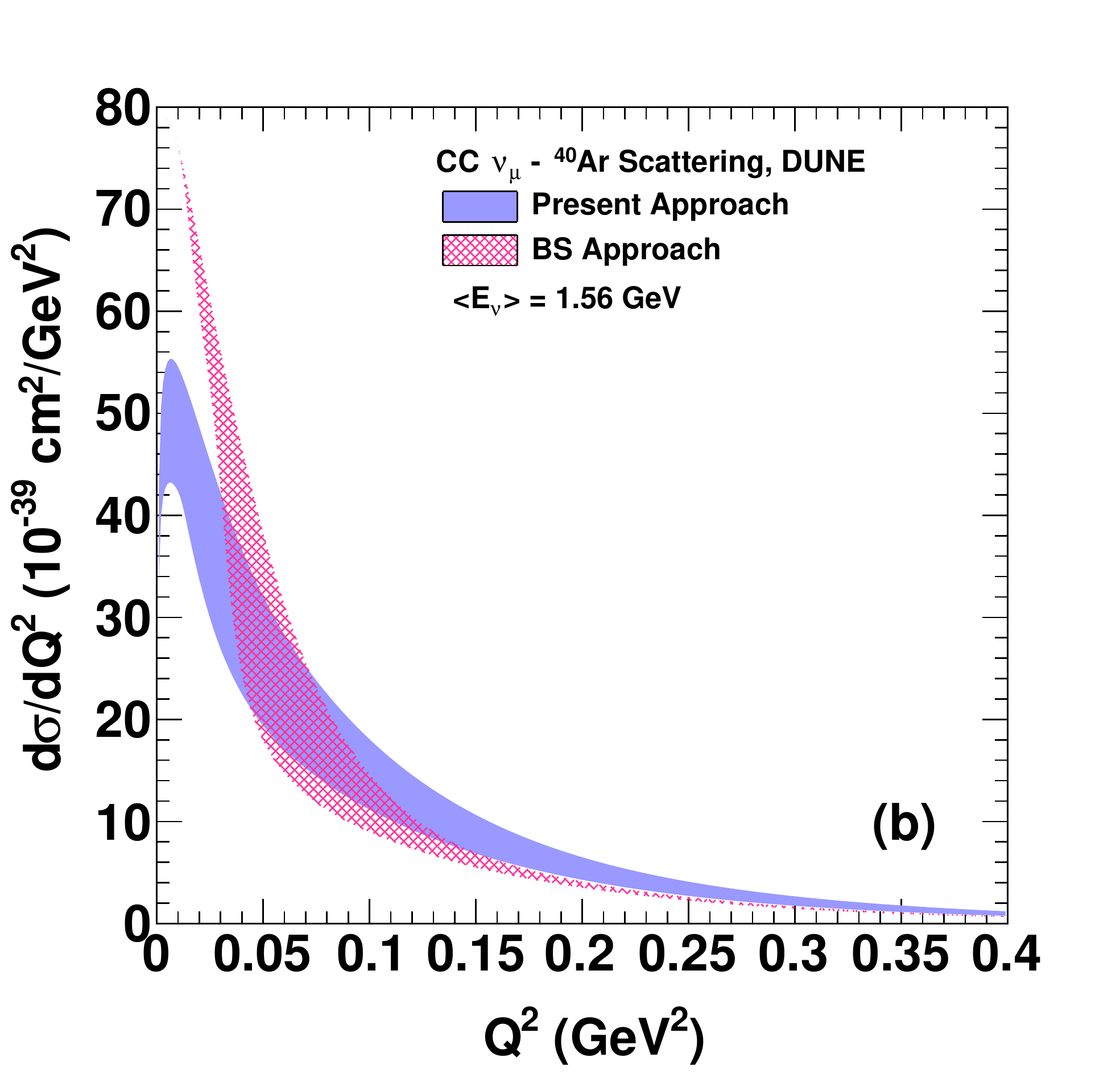}

\caption{Charge current coherent pion production in  $\nu_{\mu}-^{40}$Ar
  interaction obtained using
the Glauber model based present approach and BS approach
  (a) Total cross section
  predictions as a function of neutrino energy
  compared with GENIE (b) Differential
  cross section (averaged over neutrino flux corresponding to the DUNE
  experiment~\cite{Abi:2020qib}) as a function of the square of four-momentum transfer $Q^{2}$.
} 
\label{dune_nu_Ar_total_cc}
\end{figure*}

Figure\,\ref{dune_nu_Ar_total_cc} (a) shows the total cross section for
the charge current coherent  pion production in neutrino-argon interaction
as a function of neutrino energy obtained using
the Glauber model based present approach and BS approach
as prediction for the DUNE experiment.
The calculations are compared with GENIE.
Figure\,\ref{dune_nu_Ar_total_cc} (b) shows differential
cross section(averaged over neutrino flux
corresponding to the DUNE experiment~\cite{Abi:2020qib}) as a function of the
square of four-momentum transfer $Q^{2}$ calculated using Glauber based
approach and BS approach.

\begin{figure*}
\includegraphics[width=0.48\linewidth]{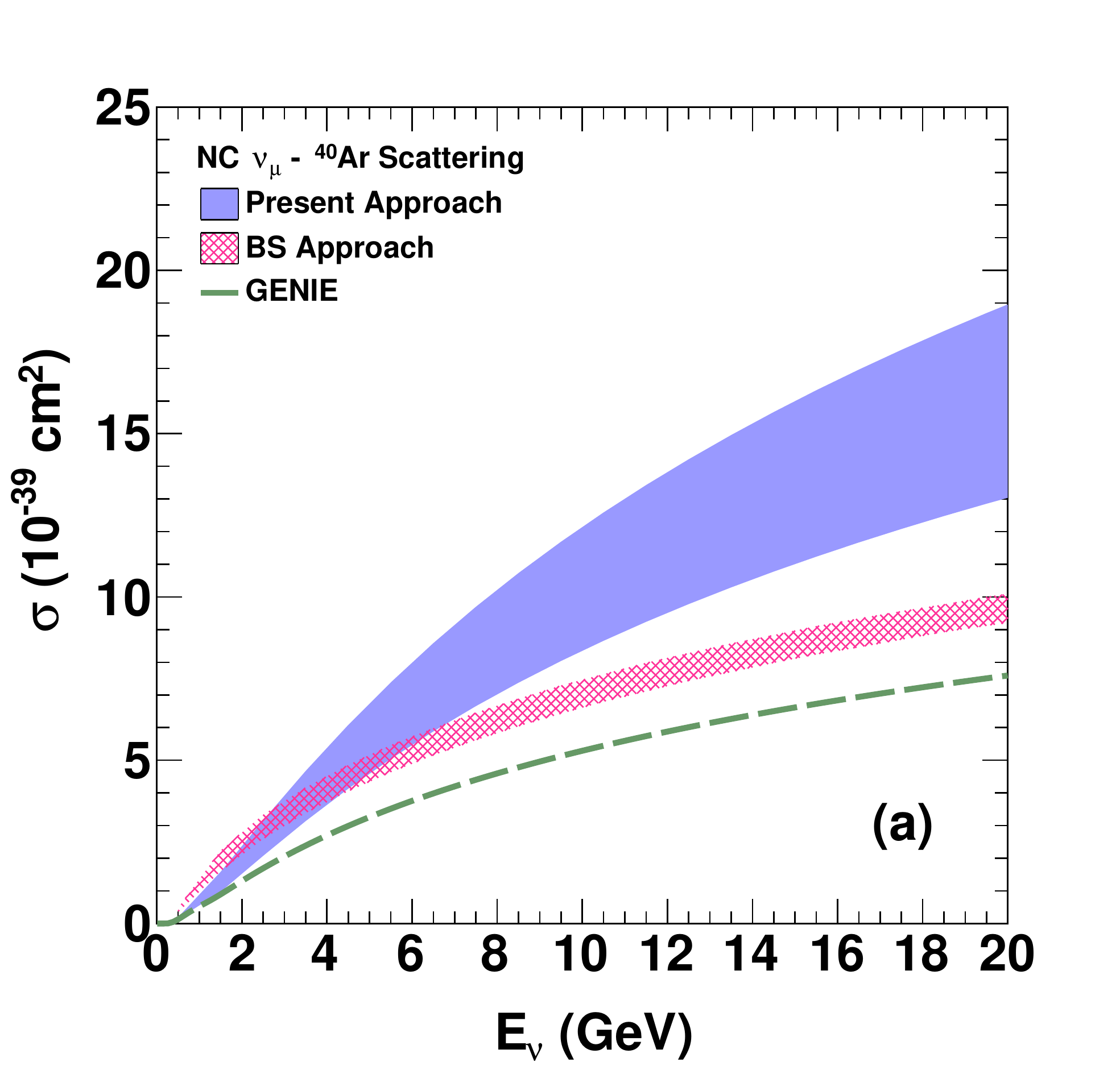}
\includegraphics[width=0.48\linewidth]{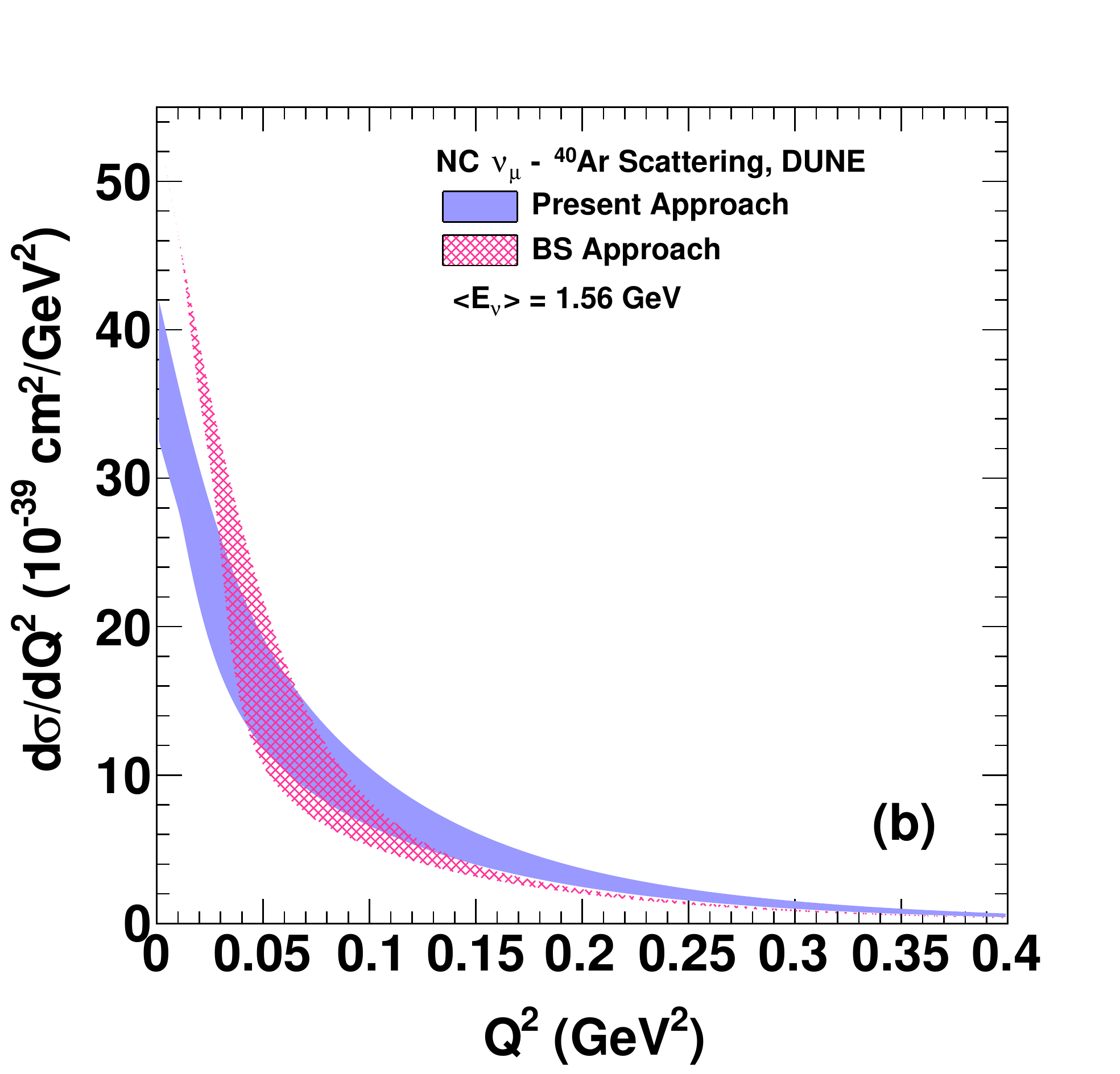}

\caption{Neutral current coherent pion production in  $\nu_{\mu}-^{40}$Ar
  interaction obtained using
the Glauber model based present approach and BS approach
  (a) Total cross section
  $\sigma$ as a function of neutrino energy are compared with GENIE (b) Differential
  cross section (averaged over neutrino flux corresponding to the DUNE
  experiment~\cite{Abi:2020qib}) as a function of the square of four-momentum transfer $Q^{2}$.
} 
\label{dune_nu_Ar_total_nc}
\end{figure*}

Figure\,\ref{dune_nu_Ar_total_nc} (a) shows the total cross section for
the neutral current coherent  pion production in neutrino-argon interaction
as a function of neutrino energy obtained using
the Glauber model based present approach and BS approach
as predictions for the DUNE experiment.
The calculations are compared with GENIE.
Figure\,\ref{dune_nu_Ar_total_nc} (b)
shows differential cross section (averaged over neutrino flux corresponding to the DUNE
experiment~\cite{Abi:2020qib}) as a function of the square of four-momentum
transfer $Q^{2}$ calculated using the Glauber model based present approach and BS approach.

\begin{figure*}
\begin{center}
\includegraphics[width=0.6\textwidth]{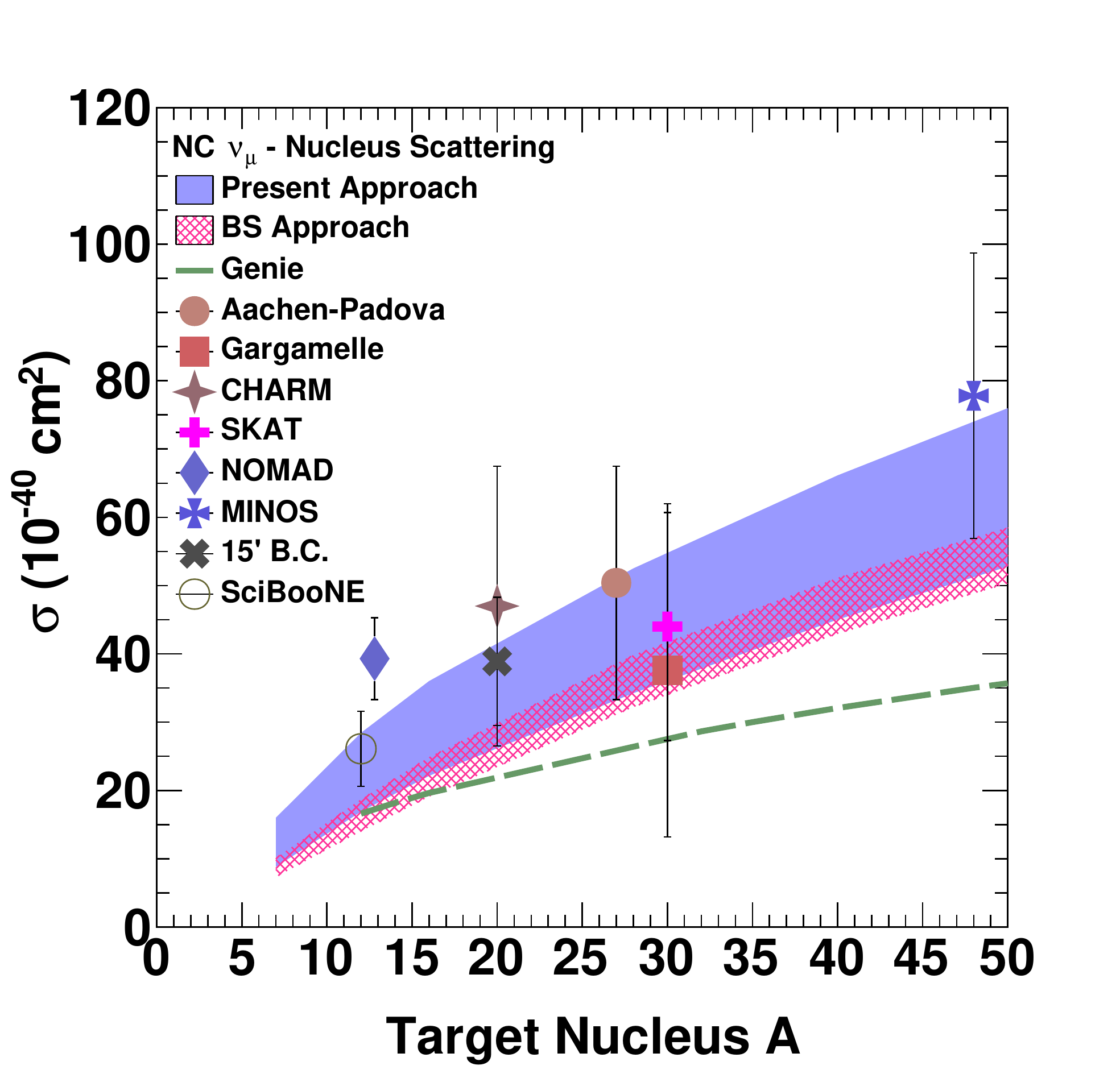}
\caption{Total cross section $\sigma$ for the neutral current coherent pion
  production in $\nu_{\mu}$-nucleus interaction as a function of nuclear mass
  $A$ at 4.9 GeV neutrino energy
  obtained using the Glauber model based present approach.
  The measurements of experiments are scaled
  to $E_{\nu}$ = 4.9 GeV~\cite{Adamson:2016hyz} using
  the Berger-Sehgal~\cite{Berger:2008xs} approach. The results obtained using
  our calculations are compared with BS approach, GENIE and experiments:
  Aachen-Padova~\citep{faissner:1983}, Gargamelle~\cite{isiksal:1984},
  CHARM~\cite{Bergsma:1985qy}, SKAT~\cite{Grabosch:1985mt},
  NOMAD~\cite{Kullenberg:2009pu}, 15'B.C.~\cite{Baltay:1986cv} and
  SciBooNE~\cite{Kurimoto:2009wq} }
\label{NeutralA}
\end{center}
\end{figure*}

Figure\,\ref{NeutralA} shows the total cross section for the neutral current
coherent pion production in  neutrino-nucleus interaction as a function
of target nucleus mass ($^{7}$Li, $^{12}$C, $^{16}$O, $^{28}$Si, $^{40}$Ar and $^{56}$Fe)
obtained using the Glauber model based present approach.
The calculations are compared with BS approach, GENIE and
experiments: Aachen-Padova~\citep{faissner:1983},
Gargamelle~\cite{isiksal:1984}, CHARM~\cite{Bergsma:1985qy},
SKAT~\cite{Grabosch:1985mt}, NOMAD~\cite{Kullenberg:2009pu},
15'B.C.~\cite{Baltay:1986cv} and SciBooNE~\cite{Kurimoto:2009wq}. The
calculations give a very good description for several experimental data for a wide
range of targets and are much better than GENIE. The measurements of experiments
are scaled to $E_{\nu}$ = 4.9 GeV~\cite{Adamson:2016hyz} using
the Berger-Sehgal~\cite{Berger:2008xs} approach.

\begin{figure*}
\begin{center}
\includegraphics[width=0.6\textwidth]{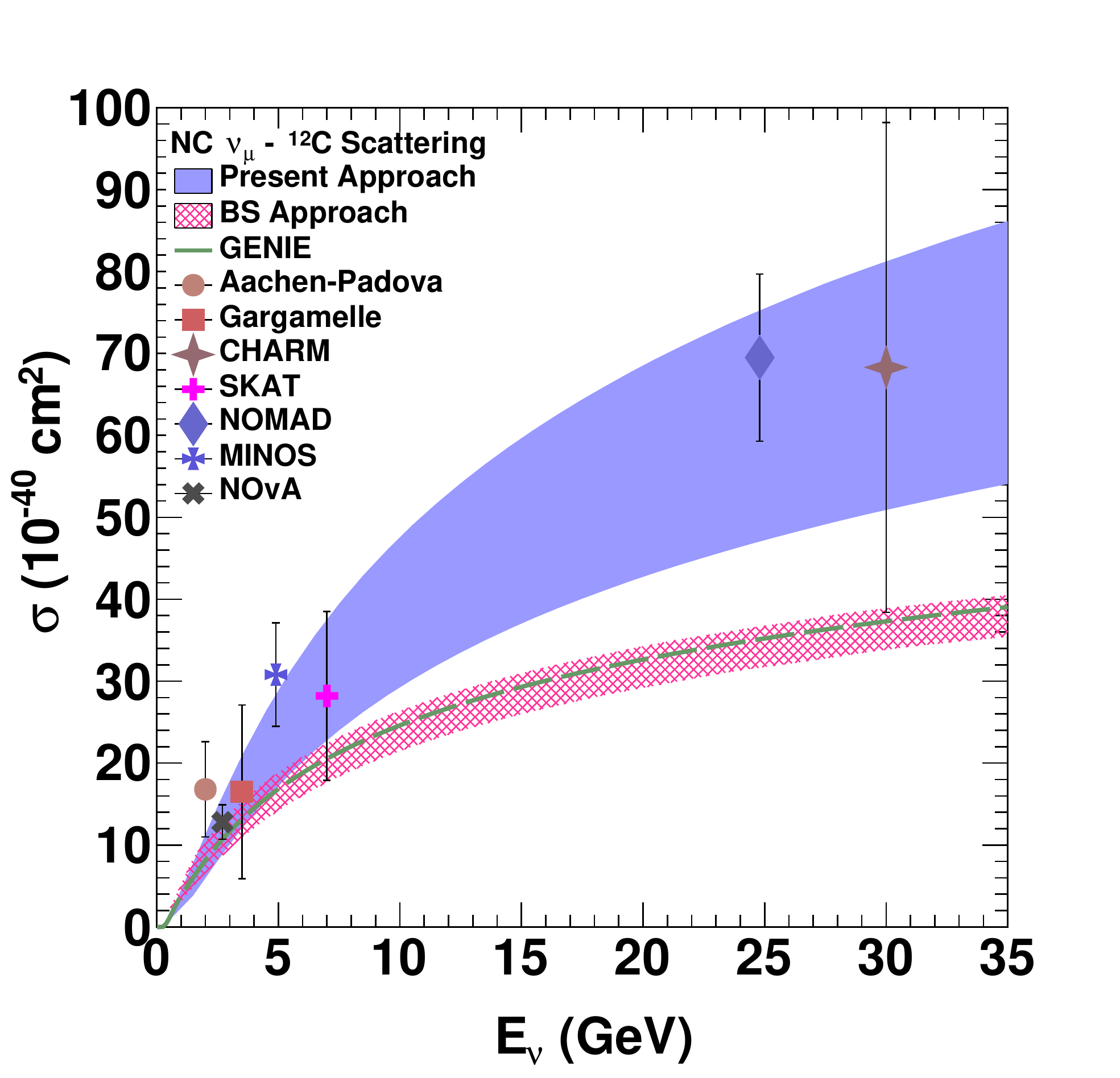}
\caption{Total cross section $\sigma$ for the neutral current coherent pion
  production in  $\nu_{\mu}- ^{12}$C interaction as a function of neutrino
  energy. The cross sections obtained using
the Glauber model based present approach and BS approach
  are compared with
  GENIE and  experiments: Aachen-Padova~\citep{faissner:1983},
  Gargamelle~\cite{isiksal:1984}, CHARM~\cite{Bergsma:1985qy},
  SKAT~\cite{Grabosch:1985mt}, NOMAD~\cite{Kullenberg:2009pu},
  MINOS~\cite{Adamson:2016hyz} and NO$\nu$A~\cite{Acero:2019qcr}.
 The cross section
 measurements of experiments with different nuclei of mass number $A$ are scaled to
 carbon~\cite{Adamson:2016hyz} using $(12/A)^{\frac{2}{3}}$.}
\label{all_exp_scaled_c12_nc}
\end{center}
\end{figure*}
Figure\,\ref{all_exp_scaled_c12_nc} shows the total cross section for
the neutral current coherent pion production in neutrino-carbon
interaction as a function of neutrino energy obtained using
the Glauber model based present approach and BS approach.
The calculations
are compared with GENIE and experiments: Aachen-Padova~\citep{faissner:1983},
Gargamelle~\cite{isiksal:1984}, CHARM~\cite{Bergsma:1985qy},
SKAT~\cite{Grabosch:1985mt}, NOMAD~\cite{Kullenberg:2009pu},
MINOS~\cite{Adamson:2016hyz} and NO$\nu$A~\cite{Acero:2019qcr}.
The Glauber based calculations give a very good description for experimental data 
in a wide range of energy. GENIE and BS approach give a good description of data only at low
neutrino energies. The cross section
measurements of experiments with different nuclei of mass number $A$ are scaled to
carbon~\cite{Adamson:2016hyz} using $(12/A)^{\frac{2}{3}}$.

\begin{table}[ht]
  \caption{Comparison of the calculated total cross sections $\sigma$ for coherent pion
    production using Glauber and BS approach with the measurements by different experiments
    along with the average energy
    of neutrinos.}
\begin{center}
\begin{tabular}{ |c|c|c|c|c|c|c|c| } 
 \hline
 Exps. & $<A>$ & Current, &$ \left\langle E \right\rangle$ & \multicolumn{3}{c|} {$\sigma$ ($10^{-39} cm^2$)}  \\
 \cline{5-7}
 &  &  particle & GeV & Glauber & BS & Exp. \\
\hline
 SKAT~\cite{Grabosch:1985mt}   & 30   & CC,   $\nu$ &   7    & 8.8-13.3 & 7.7-9.2 & 10.6$\pm$1.6 \\ 
 SKAT~\cite{Grabosch:1985mt}   & 30   & CC,   $\bar{\nu}$ &7 & 8.8-13.3 & 7.7-9.2 & 11.3$\pm$3.5 \\  
 SKAT~\cite{Grabosch:1985mt}   & 30   & NC,   $\nu$ &   7    & 5.0-7.5  & 4.3-5.1 & 5.2$\pm$1.9 \\   
 MINOS~\cite{Adamson:2016hyz}  & 48   & NC,   $\nu$ &   4.9  & 5.1-7.3  & 4.9-5.7 & 3.26$\pm$0.21 \\
 NO$\nu$A~\cite{Acero:2019qcr} & 13.8 & NC,   $\nu$ &   2.7  & 1.0-1.8  & 1.0-1.5 & 1.4$\pm$0.2 \\ 
 \hline
\end{tabular}
\end{center}
\label{table2}
\end{table}

Table~\ref{table2} shows a comparison of the calculated total cross sections $\sigma$
for coherent pion production using Glauber and BS approach
with the measurements by different experiments SKAT~\cite{Grabosch:1985mt},
MINOS~\cite{Adamson:2016hyz} and NO$\nu$A~\cite{Acero:2019qcr} along with the
average energy of neutrinos.
In the SKAT experiment, target material is Heavy freon ($A$ = 30) and the calculation is done
on the average energy of neutrinos in the experiment.
In the MINOS experiment, target material is the composition of 80$\%$ iron and 20$\%$
carbon ($A$ = 48). The cross section is averaged over the flux.
In the NO$\nu$A experiment target material is mainly the composition of 66.7$\%$ carbon,
16.1$\%$ chlorine and 10.8$\%$ hydrogen and other nuclei ($A$ = 13.8).
The table shows excellent agreement between the data and the calculations, especially
at higher energy.

\section{\bf Conclusions}

  In this work, we presented a comprehensive study of coherent pion production in neutrino-nucleus
interactions in the resonance region using the formalism based on PCAC theorem.
Pion-nucleus elastic scattering cross section is calculated using the Glauber
model which takes three inputs, nuclear densities, pion-nucleon cross section and
$\alpha_{\pi N}$ for which the parametrizations are obtained from measured data.
We obtain the differential and integrated cross sections for charge and
neutral current coherent pion production in neutrino (anti-neutrino)-nucleus
scattering for a range of targets such  as lithium,  carbon, hydrocarbon, scintillator,
oxygen, silicon, argon, iron and lead. The results of these cross section calculations
are compared with the
measured data, BS model and GENIE package. There is an excellent agreement between the calculated
cross section fom Glauber model and measured cross sections. Predictions are also
made for upcoming experiments like INO and DUNE in the coherent pion production region
of the neutrino cross section.

{\bf Acknowledgements}
 We thank Dr. Kapil Saraswat for many fruitful discussions during the course of this
work.

\end{document}